\documentclass{aa}  

\usepackage{graphicx}
\usepackage{xfrac}
\usepackage[version=4]{mhchem}
\usepackage{txfonts}
\usepackage{float}
\usepackage{subcaption}
\usepackage[colorlinks=true,urlcolor=blue,citecolor=blue,linkcolor=blue,breaklinks=true]{hyperref}
\newcommand{\subsubsubsection}[1]{\paragraph{#1}\mbox{}\\}
\begin{document}

   \title{Circumplanetary disk ices}

   \subtitle{I. Ice formation vs. viscous evolution and grain drift}

   \author{N. Oberg
          \inst{1,2}
          \and
          I. Kamp 
          \inst{1} 
          \and
          S. Cazaux
          \inst{2,3}
          \and 
          P. Woitke
          \inst{4,5}
          \and 
          W.F.Thi
          \inst{6}
          }

   \institute{Kapteyn Astronomical Institute, University of Groningen, P.O. Box 800, 9700 AV Groningen, The Netherlands \\
              \email{oberg@astro.rug.nl}
         \and
             Faculty of Aerospace Engineering, Delft University of Technology, Delft, The Netherlands 
         \and
             University of Leiden, P.O. Box 9513, 2300 RA, Leiden, The Netherlands   
         \and 
             Space Research Institute, Austrian Academy of Sciences, Schmiedlstrasse 6, A-8042 Graz, Austria
         \and 
             Centre for Exoplanet Science, University of St. Andrews, North Haugh, St. Andrews, KY16 9SS,UK   
         \and 
             Max-Planck-Institut für extraterrestrische Physik, Giessenbachstrasse 1, 85748 Garching, Germany \\
             }

   \date{Received --- accepted ---}

% \abstract{}{}{}{}{} 
% 5 {} token are mandatory
 
  \abstract
   {The large icy moons of Jupiter formed in a circumplanetary disk (CPD). CPDs are fed by vertically infalling circumstellar gas and dust which may be shock-heated upon accretion. Accreted material is then either incorporated into moons, falls into the planet, or is lost beyond the disk edge on relatively short timescales.   If ices are sublimated during accretion onto the CPD we know there must be sufficient time for them to recondense or moons such as Ganymede or Callisto could not form. The chemical timescale to form sufficiently icy solids places a novel constraint on the dynamical behaviour and properties of CPDs.}
   {We aim to explore the process of ice formation in CPDs to constrain which disk properties (such as the mass, viscosity, and dust-to-gas ratio) are consistent with the formation of an icy moon system.}
   {We use the radiation thermochemical code \textsc{ProDiMo} (Protoplanetary Disk Model) to analyze how the radial ice abundance evolves in CPDs.  We consider different initial chemical conditions of the disk to explore the consequences of infalling material being inherited from the circumstellar disk or being reset to atomic conditions by shock-heating. We contrast the timescales of ice formation with disk viscous timescales and radial dust drift.}
   {We have derived the radial ice abundance and rate of ice formation in a small grid of model CPDs. Water ice can form very efficiently in the CPD from initially atomic conditions, as a significant fraction is efficiently re-deposited on dust grains within < 1 yr. Radial grain drift timescales are in general longer than those of ice formation on grains. Icy grains of size $a < 3$ mm retain their icy mantles while crossing an optically thin circumstellar disk gap at 5 au for $L_* < 10 $ L$_{\odot}$.}
   {Three-body reactions play an important role in water formation in the dense midplane condition of CPDs.  The CPD midplane must be depleted in dust relative to the circumstellar disk by a factor 10-50 to produce solids with the ice to rock ratio of the icy Galilean satellites. The CPD snowline is not erased by radial grain drift, which is consistent with the compositional gradient of the Galilean satellites being primordial.}


   \maketitle
%
%-------------------------------------------------------------------

\section{Introduction}

A general feature of regular satellite formation theory is that the circumplanetary disk (CPD) consists of circumstellar material accreted from within the vicinity of the planet \citep{Lubow1999, Canup2002, Alibert2005,Shibaike2019, Ronnet2020}.  If the planet is massive enough to open a gap in the circumstellar disk, material continues to flow into the gap \citep{Kley1999, Teague2019} and falls nearly vertically onto the CPD  \citep{Tanigawa2012, Morbi2014}.  The CPD achieves a steady-state mass when this inflow is balanced by outflow where gas either spirals into the planet or is decreted beyond the Hill sphere \citep{Canup2002,Batygin2020}.  Independently of disk viscosity, stellar tides induce spiral waves in the CPD which transport angular momentum and promote accretion onto the planet at a rate on the order $10^{-7}$\,M$_{\rm J}$ yr$^{-1}$  \citep{Rivier2012}, suggesting that for a CPD of mass $\sim10^{-4}$ M$_{\rm J}$ \citep{2002A&A...385..647D,Gressel2013,2014ApJ...782...65S} infalling gas spends only a limited time inside the CPD before being lost \citep{Canup2002}.  

The timescale of radial dust drift in small disks is also predicted to be short \citep{Pinilla2013,Shibaike2017,Rab2019}. A CPD could lose mm-size grains within \mbox{10$^2$-10$^3$}\,yr to aerodynamic drag against highly sub-Keplerian gas due to a very steep radial pressure gradient \citep{Zhu2018}.  The CPD is thus a very dynamical system, potentially with both inwards and outwards radial transport of both gas and dust on very short timescales.   

The amount of time which gas and dust spend within the CPD becomes highly relevant if a chemical ``reset" occurs. The infalling circumstellar material may pass through one or more accretion shocks \citep{Lubow1999,Tanigawa2012,Zhu2015,Szulagyi2016,Schulik2020} and can be heated $\geq 1000$\,K during accretion onto the CPD \citep{Szulagyi2017,Szulagyi2017b,Aoyama2018}.  For pre-shock velocities in excess of 90\,km\,s$^{-1}$ the shock can be sufficiently hot to leave most of the infalling gas atomic or ionized \citep{Aoyama2018}. The shock may also desorb the icy mantles of grains via sputtering and thermal desorption, which for pre-shock velocities in excess of 8-10\,km\,s$^{-1}$ can effectively strip a grain of H$_2$O ice \citep{Woitke1993,Aota2015,tielens2021}. Small icy grains passing through the gap may also lose their volatile contents to photodesorption prior to shock-heating \citep{Turner2012}.  We refer to a "reset" scenario if shock-heating or photodesorption effectively destroys molecules in the accretion flow to the CPD.  In a reset scenario, the re-formation of ices in the CPD must compete with viscous accretion and decretion of gas and radial drift of dust. Alternatively, if circumstellar disk ices survive incorporation into the CPD, we refer to an ``inheritance" scenario.   

The Galilean satellites characteristically exhibit a radial compositional gradient of decreasing density with increasing distance from Jupiter. The inner moon Io is ice-free, while Europa has an ice mass fraction of $\sim 6$-9$\%$ \citep{schubert2004, Kuskov2005} and Ganymede and Callisto have ice mass  fractions of 40-55$\%$ \citep{mckinnon1999,schubert2004,Sohl2002}.  Theoretically it appears challenging to reproduce the compositional gradient by tidal heating \citep{Bierson2021} or impact-driven escape of volatiles \citep{dwyer2013}.  Previously it has been proposed that the gradient was imprinted during the formation of the moons by a radial temperature gradient in the CPD \citep{Lunine1982}, but the relevant chemical timescales have rarely been taken into account \citep{Mousis2006}. It is an open question whether the gradient can be produced primordially if the chemistry of infalling gas and dust is reset.  By analyzing the composition and abundance of ices that are able to form within the relevant timescales we can place a lower bound on how efficiently angular momentum is transported within the CPD.  

In this work we investigated the balance of the competing timescales of ice formation, dust grain drift, and gas viscous flow to seek constraints on properties of the CPD such as viscosity, ice mass fraction, and dust-to-gas ratio.  We considered the two opposing extreme cases of full chemical inheritance and reset in chemically evolving disk models utilizing a rate-based modeling approach. In Sect. \ref{sec:methods} we describe our modeling set-up and the assumptions that we make. In Sect.  \ref{sec:results} we analyze the CPD time-dependent ice abundances for both the reset and inheritance cases, and place novel constraints on the properties of CPDs.  In Sect. \ref{sec:discussion} we discuss the implications of the these constraints and place them in the context of solar system formation.  We consider also the role that radial grain drift has in competing with ice adsorption and desorption. In Sect. \ref{sec:conclusions} we summarize our key findings.

\section{Methods} \label{sec:methods}

We considered two opposing scenarios; one in which the molecular gas-phase and ice chemistry of the circumstellar disk is preserved during accretion onto the CPD (inherit) and one in which it is lost (reset). In the former case the CPD is initially populated with gas and ices extracted from the circumstellar disk. In the latter case the disk is initially populated by a fully atomic gas and dust is free of ice. We followed the build-up of ices in the CPD over time in a thermochemical disk model (see Sect. \ref{sec:methods:diskmodeling}) and extracted the molecular ice abundance and composition as a function of time.

The net inflow and outflow rate of gas and solids from and to the CPD is assumed to be zero, and that the disk mass is in steady-state. The rate of gas outflow then sets an upper limit on the applicable chemical evolutionary timescales.  Hereafter we refer to this as the viscous timescale, defined as

\begin{equation} \label{eq:tvisc}
    t_{\rm visc} = \frac{M_{\rm CPD}}{\dot M} ,
\end{equation}

\noindent
where $M_{\rm CPD}$ is the mass of the CPD and $\dot M$ is the infall rate of circumstellar material onto the CPD.  The relatively short $t_{\rm visc}$ considered in this work (see Sect. \ref{sec:cpdviscosity}) may cause reactions with high activation energies to be kinetically inhibited, although it has been noted that the relatively high densities characteristic of CPDs may allow these reactions to proceed to equilibrium \citep{dePater2010}. Nevertheless we contrast the time-dependent results with the assumption of steady-state chemistry. In steady-state chemistry the disk is allowed to evolve for an indefinite time period until the rate of formation for every gas and ice species is balanced by the corresponding rate of destruction.  

\subsection{Disk modeling code} \label{sec:methods:diskmodeling}

We used the radiation thermochemical disk modeling code \textsc{ProDiMo} \footnote{https://www.astro.rug.nl/~prodimo/} (\textbf{Pro}toplanetary \textbf{Di}sk \textbf{Mo}del)  \citep{Woitke2009a,Woitke2016,Kamp2010,Kamp2017,Thi2011,Thi2018H2,Thi2020H2} to explore the formation rates and resulting abundances of various ices in CPDs. \textsc{ProDiMo} uses a rate equation based approach to compute the gas chemistry using either a time-dependent or steady-state solver.  The model represents a 2D slice through an axisymmetric disk, extending radially in distance from the planet $r$ and vertically in distance from the disk midplane $z$.   Our chemical network contains 13 elements and 235 atomic and molecular species. Where not explicitly specified we used the ``large DIANA chemical standard" network as described in \citet{Kamp2017}.  Reaction rates are mainly selected from the UMIST2012 database \citep{McElroy2013}. Important three-body collider reactions are adopted from the UMIST2006 rate file \citep{Woodall2007}.  Gas-phase reactions within the CPD produce molecular species which can then freeze-out on grain surfaces.  The rate of ice formation is determined by the available grain surface area, dust temperature, and the rates of thermal, photo-, and cosmic-ray desorption (see Sect. \ref{sec:iceform} for a detailed description of ice formation).  

We made the simplifying assumption that the chemical composition of the infalling material is distributed instantaneously and homogeneously throughout the disk (see appendix \ref{appendix:vertical-mixing} for a discussion of the potential impact of the rate of vertical gas mixing). We assumed that the CPD inherits micrometer to mm-sized dust grains directly from the circumstellar disk. The vertical dust stratification is calculated according to the method of \citet{Dubrulle1995} and kept fixed prior to the calculation of chemical evolution. The timescales of radial dust drift are calculated in a post-processing step described in \ref{sec:methods:drift}.    The temperature and radiation structure of the CPD is solved in steady-state and then kept fixed during chemical evolution of inherited or reset infalling material.

In Sect. \ref{sec:results_surface} we considered the implications of including grain-surface chemistry reactions.  With surface chemistry \textsc{ProDiMo} models explicitly the formation of H$_2$ in the CPD for which the inclusion of additional chemical species such as hydrogenated PAH is necessitated \citep{Thi2020,Thi2020H2}.  The selection of the additional species and reactions in the surface chemistry network and their role in the eventual composition of the ices will be discussed in an accompanying work focused on the composition of the ices (Oberg et al. in prep.).

\subsubsection{Ice formation} \label{sec:iceform}

Where conditions in the disk are appropriate, ices can condense onto the grains in successive layers by physical van der Waals bonding (physisorption). The adsorption rate of species $i$ onto a physisorption sites is 

\begin{equation}
  R_{i}^{\rm ads} = 4 \pi a^2 v_{i}^{\rm th} n_{\rm d} S_{i} \, \, \rm  s^{-1} ,
\end{equation}
\noindent 
where 4$\pi a^2 $ is a dust grain surface area, $a$ is the grain radius, v$_{i}^{\rm th}$ is the thermal speed ($k_{\rm B} T_{\rm g} / 2 \pi m_i)^{1/2}$, $k_{\rm B}$ is the Boltzmann constant, $T_{\rm g}$ is the gas temperature, $m_{i}$ is the mass of the gas-phase species, $n_{\rm d}$ is the dust number density, and  $S_{i}$ is the sticking coefficient. The ice adsorption rate coefficients are then
\begin{equation}
    \frac{dn_{\#,i}}{dt} = R_{i}^{\rm ads} n_i \, \, \rm cm^{-3}  \rm s^{-1} ,
\end{equation}
\noindent
where $n_i$ is the number density of the gas-phase species. 

Physisorbed species can desorb thermally, photodesorb, or desorb after a cosmic ray impact deposits energy in the grain. For physisorption there is no desorption barrier and the desorption energy is equal to the binding energy E$_{i}^{\mathrm{b}}$.  The Arrhenius formulation for the rate of thermal desorption is then

\begin{equation}
R_{i}^{\rm des,\mathrm{th}}=v_{0, i} e^{-E_{i}^{\mathrm{b}} / k_B T_{\mathrm{d}}} \text { in s }^{-1} \text {, }
\end{equation}

\noindent
where the pre-exponential (frequency) factor $v_{\rm 0, i}$ is

\begin{equation}
v_{0, i}=\sqrt{\frac{2 N_{\mathrm{surf}} E_{i}^{\mathrm{b}}}{\pi^{2} m_{i}}}.
\end{equation}
\noindent
 $N_{\rm surf}$ is the density of surface binding sites 1.5$\times10^{15}$ cm$^{-2}$ \citep{Hasegawa1992}, and $T_{\rm d}$ is the dust temperature. Adsorption energies of major volatile species are listed in Appendix \ref{appendix:eads}. 
 
The photodesorption rate is computed using the UV field strength calculated by the radiative transfer and photodissociation cross-sections.  The photodesorption rate for species $i$ is 
 
 \begin{equation}
     R_{i}^{\rm des,ph} = \pi a^2 \frac{n_{\rm d}}{n^{\rm act}_i} Y_i \, \chi F_{\rm Draine} \, \textrm{s}^{-1} ,
 \end{equation}
 
\noindent
where $Y_{i}$ is the photodesorption yield, $n^{\rm act}_i$ is the concentration of the species in the active layers,  
 
\begin{equation}
\begin{aligned}
n_{i}^{\text {act }} &=n_{i} &  \text { if } n_{\#, \text { tot }} \leqslant 4 \pi N_{\rm surf} a^2 n_{\mathrm{d}} \\
&=n_{i}\left(N_{\text {act }} / N_{\text {layer }}\right) & \text { if } n_{\#, \text { tot }}> 4 \pi N_{\rm surf} a^2 n_{\mathrm{d}} .
\end{aligned}
\end{equation}
\noindent
The number of physisorbed layers $N_{\rm layer}$ = $
n_{\#, \text { tot }} / (4 \pi N_{\rm surf} a^2 n_{\mathrm{d}}) $ and   $ n_{\#, \text { tot }}=\sum_{i} n_{\#, i}$ is the total number of the physisorbed species, $4\pi N_{\rm surf}a^2$ is the number of binding sites per layer, and $N_{\rm act}$ is the number of chemically active layers. $\chi F_{\rm Draine}$ is a measure of the local UV energy density from the 2D continuum radiative transfer \citep{Woitke2009a}. The rate of cosmic-ray induced desorption $R_{i}^{\rm des,CR}$ is calculated according to the method of \citep{Hasegawa1993}.  The total desorption rate $R_{i}^{\rm des}$ is the sum of the thermal, photo- and cosmic ray induced desorption rates $R_{i}^{\rm des,th}$+$R_{i}^{\rm des,ph}$+$R_{i}^{\rm des,CR}$. The desorption rate of the physisorbed species $i$ is then

\begin{equation}
    \frac{dn_i}{dt} = R_{i}^{\mathrm{des}} n_{i}^{\rm act} \, \textrm{cm}^{-3} \textrm{s}^{-1} ,
\end{equation}
\noindent

\begin{figure}
  \includegraphics[width=\textwidth/2]{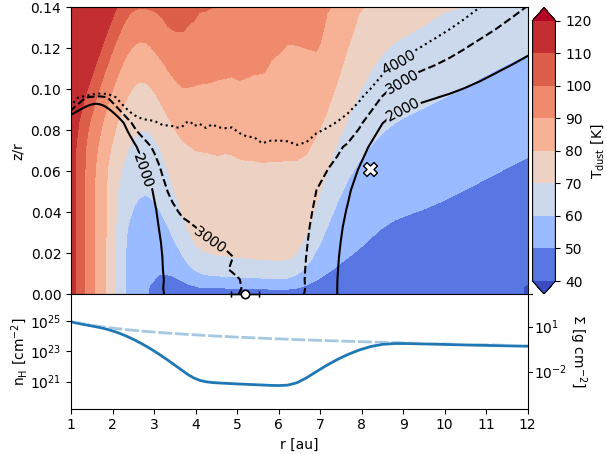}
      \caption{  2D dust temperature distribution around the circumstellar disk gap in the range 40-120\,K (\textit{top}).  Solid, dashed, and dotted black contours indicate the relative UV field strength $\chi$ (see Eq.\ref{eq:chi}).  The white circle with black border represents the position of Jupiter and the horizontal bar indicates the physical extent of the CPD.  The white cross with black border represents the location from which the inherited chemistry is extracted at one scale height $z = H$.  Hydrogen nuclei column density (blue line) with unperturbed circumstellar disk density profile for comparison (light blue dashed line) (\textit{bottom}).}
      \label{fig:mmsn-gap-conditions}
\end{figure}

\subsubsection{Properties of the Disk Models}

\begin{table}

    \caption{Parameters for the solar circumstellar disk.}
    
    \centering
    \renewcommand{\arraystretch}{1.1}%

   \begin{tabular}{llll}
    \hline \hline
        Parameter                 & Symbol             & Value        & Unit          \\ \hline
        Stellar Mass              & $M_*$              & 1.0          &  M$_{\odot}$  \\
        Stellar Luminosity        & $L_*$              & 0.84         & L$_{\odot}$   \\
        
        Effective Temperature     & $T_{\rm eff}$      & 4395         & K             \\
        UV Luminosity             & $L_{\rm UV,*}$     & 0.01         &  L$_{\odot}$  \\
        X-ray Luminosity          & $L_{\rm X}$        & 10$^{30}$    & erg s$^{-1}$  \\
  
        \hline  
  
        Disk Mass                 & $M_{\rm disk}$     & 0.001       &  M$_{\odot}$ \\
        Disk Inner Radius         & $R_{\rm in} $      & 0.1         & au           \\ 
        Disk Outer Radius         & $R_{\rm out} $     & 100         & au           \\
        Column Density Power Ind. & $\epsilon$         & 1.5         & -            \\
        Reference Scale Height    & H$_{\rm 10 au}$    & 1           & au           \\
        Flaring Index             & $\beta$            & 1.15        & -            \\ \hline
        
        Minimum dust size         & $a_{\rm min}$      & 0.05        & \textmu m \\
        Maximum dust size         & $a_{\rm max}$      & 3000        & \textmu m \\
        Dust size power law index & $a_{\rm pow}$      & 3.5         & -         \\
        Dust-to-Gas ratio         & $d/g$              & $10^{-2}$   & -         \\ \hline
        
        Dust composition: \\
         \hspace{0.5cm} Mg$_{0.7}$Fe$_{0.3}$SiO$_3$ &    & $60\%$ \\
         \hspace{0.5cm} Amorphous carbon            &    & $15\%$ \\
         \hspace{0.5cm} Vacuum                      &    & $25\%$ \\

    \end{tabular}
\caption*{\textbf{Note}: Stellar temperature and luminosity are selected from the pre-main sequence stellar evolutionary tracks of \citet{2000A&A...358..593S} for $t = 4$ Myr. Stellar UV and X-ray luminosities for a representative Class II T Tauri star are adopted from \citet{2016A&A...586A.103W}.  }
    \label{tab:ppds}
\end{table}

\subsubsubsection{The Circumstellar Disk}

To generate the chemical abundances for our ``inheritance" scenario we considered the properties of the surrounding circumstellar disk from which material is accreted onto the CPD.  We used a two-step approach to model the chemistry in our circumstellar disk.  As a first step, the initial conditions are derived from a zero-dimensional ``molecular cloud" model, the parameters of which are listed in Table \ref{tab:mol_cloud}.  This stage represents 1.7$\times10^5$\,yr (the estimated lifetime of the Taurus Molecular Cloud TMC-1) of chemical evolution in a pre-collapse molecular cloud state \citep{McElroy2013}. The resulting chemical abundances of the majority of the most common species agree within a factor 10 with the observed abundances in TMC-1(see Fig. \ref{fig:appendix-mc}). These abundances are then used as initial conditions for the 2D grid of cells in the circumstellar disk model in the second step. 

In the second step the circumstellar disk model is evolved for an additional 4\,Myr to be consistent with the formation timeline of Jupiter proposed to account for the distinct isotopic population of meteorites found in the solar system wherein Jupiter undergoes runaway accretion $>$3.46\,Myr after the formation of calcium-aluminium rich refractory inclusions \citep{Kruijer, Weiss2021}. The surface density power law and physical extent of the circumstellar disk is based on a modified ``Minimum Mass Solar Nebula" (MMSN)  \citep{1981PThPS..70...35H}. A parameteric gap has been introduced which reduces the dust and gas density at 5.2\,au centered on the location of Jupiter. The gap dimensions are parameterized by an analytical gap scaling relation derived from hydrodynamical simulations  and are consistent with a circumstellar disk viscosity of $\alpha\sim10^{-4}$ and disk age of 4\,Myr \citep{2016PASJ...68...43K}.  A detailed description of the circumstellar disk model and gap structure methodology can be found in \citep{Oberg2020}. The relevant circumstellar disk model parameters can be found in Table \ref{tab:ppds}.  The dust temperature, surface density profile, and UV field strength in and around the circumstellar disk gap can be seen in Fig. \ref{fig:mmsn-gap-conditions}.

Finally we extract the chemical abundances from the circumstellar disk model at the outer edge of the gap at a radius of 8.2\,au. The gap edge is defined as the radius at which the perturbed surface density profile reaches 50$\%$ of the unperturbed profile. As material flows into the gap from above one pressure scale height \citep{Morbidelli2014}, we extract our initial conditions for the CPD model at \mbox{$z$ = $H$} (\mbox{$z$ =  0.5\,au at $r = 8.2$\,au}). The ambient conditions at this point are listed in Table \ref{tab:inherit_point}. The extracted abundances are then used as the initial chemical composition for our ``inheritance" scenario CPD.  

Throughout this work we quantify the iciness of solids with the ice mass fraction, 

\begin{equation}
    f_{\rm ice} = \frac{m_{\rm ice}}{m_{\rm rock} + m_{\rm ice}},
\end{equation}
\noindent
where $m_{\rm ice}$ is the ice mass and $m_{\rm rock}$ is the total rock (in this case, dust) mass. The $f_{\rm ice}$ of solids in the inherited circumstellar material is 0.48.  At a single pressure scale height ($z$ = $H$) there is a factor $\sim$20 reduction of the initial, global dust-to-gas ratio due to settling which is calculated according to the method of \citet{Dubrulle1995} with $\alpha_{\rm settle} = 10^{-2}$ such that the dust-to-gas ratio $d/g$ at one scale height \mbox{d/g$_{z = H}$ = 10$^{-3.2}$}.  Nevertheless we test a range of different $d/g$ values for the CPDs both above and below this value ($10^{-4}$-$10^{-2}$).

\begin{table}[] 
\caption{Conditions at the gap edge "inheritance" point of the circumstellar disk at heliocentric distance 8.2\,au, altitude 0.5\,au (1 pressure scale height) above the midplane.}
\centering
\setlength{\tabcolsep}{1pt}
\begin{tabular}{llll}
 \hline  \hline
Parameter          & Symbol            &  Value        & Unit        \\  \hline
Hydrogen density   & $n_{\rm H,in}$    &  $10^{10}$    & cm$^{-3}$   \\  
Optical Extinction & A$_{\rm V,in}$    &  1.01         & -           \\  
Dust-to-gas ratio  & $d/g\,_{\rm in}$  &  $10^{-3.23}$ & -           \\
Dust Temperature   & $T_{\rm d,in}$    &  57.0         & K           \\
Gas Temperature    & $T_{\rm g,in}$    &  57.3         & K           \\
Ice Mass Fraction  & f$_{\rm ice, in}$     &  0.48         & -           \\

\end{tabular}
\label{tab:inherit_point}
\end{table}

\subsubsubsection{Survival of icy grains passing through the gap}

\begin{figure}
  \includegraphics[width=\textwidth/2]{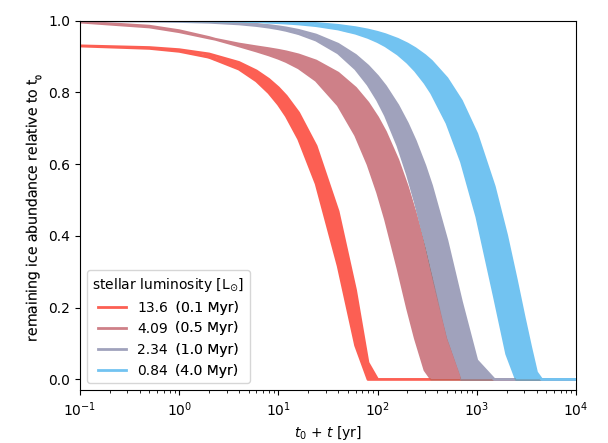}
      \caption{Ice abundance in the circumstellar disk gap normalized to the initial abundance of ice for various stellar luminosities as function of time after the onset of UV exposure.  The width of each track represents the range of ice sublimation rates corresponding to variable conditions across  the gap region ($z=0-0.5$\,au, $r=$ 5.2 - 8.2\,au).}
      \label{fig:MMSN_GAP_ICE_LOSS}
\end{figure}

Icy grains must orbit within the optically thin gap for an unknown amount of time prior to being accreted onto the CPD.  We considered whether ices on grains can survive against thermal- or photodesorption while crossing through the circumstellar disk gap. The dust and gas temperatures in the gap are not closely coupled as a consequence of the low densities. While the gas temperature extracted from the model at the midplane is $\sim$200\,K, the corresponding dust temperature is 48$\pm2$\,K.  Given that pressures in the gap range from 10$^{-12}$-10$^{-10}$\,bar, water ice is stable on the relevant timescales in the absence of irradiation \citep{Lodders2003}.  However,  the actual ice abundance at the gap midplane is negligible in steady-state. Despite the presence of a shadowing inner disk, ice within the gap are sublimated as a result of the significant stellar background radiation scattered into the gap.  %At Jupiter's location, we find a maximum far ultraviolent (FUV) $\chi$ field strength of $\sim$3000 as a result of scattered solar radiation.  

To assess the longevity of ices crossing through the gap we populated the gap region with the ``inheritance" chemical abundances found exterior to the gap and produce snapshots at regular intervals.  The resulting decline in ice abundance as a function of time is shown in Fig. \ref{fig:MMSN_GAP_ICE_LOSS} for various stellar luminosities.  The differing luminosities correspond to expected properties of the sun for a stellar age of 0.1, 0.5, 1 and 4\,Myr \citep{Siess2000}.  \citet{Turner2012} suggests that grains entering the gap are accreted generally within a single orbital period or 10\,yr.  We find that for a moon formation time \mbox{> 1 Myr} after CAI formation \mbox{(L$_{*} \leq 2.34 $L$_{\odot}$)}, grains retain $>99\%$ of their volatile content during gap-crossing if they reach the vicinity of the planet within 10 yr.  A "full" inheritance scenario is thus not excluded by conditions within the gap, but would instead rely on shock-heating at the CPD surface.

\subsubsubsection{The circumplanetary disks}

The CPD is an actively-fed accretion disk with a steady-state mass proportional to its viscosity and mass infall rate. We considered an optically thick CPD of mass $10^{-7}$ M$_{\odot}$ as well as a lower mass CPD ($10^{-8}$ M$_{\odot}$) which is optically thin everywhere outside the orbit of Callisto (\mbox{$r$ > 0.03 R$_{\rm H}$} where \mbox{R$_{\rm H}$ =  0.34 au} is the Hill radius).  For a Jovian-mass planet these represent planet-disk mass ratios of $\sim10^{-4}$ and $10^{-5}$ respectively.  The CPDs are thus of the "gas-starved" type, and do not instantaneously contain the mass required to form a moon system as massive as the Galilean one.

\begin{figure}
  \includegraphics[width=\textwidth/2]{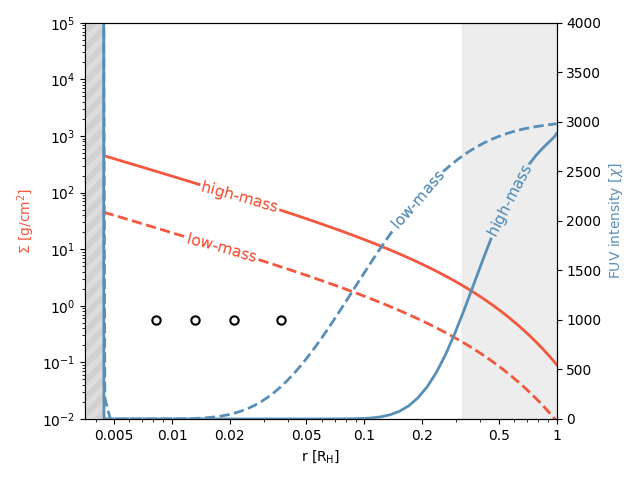}
      \caption{Surface density (red) and midplane FUV field strength in units of the Draine field $\chi$ (blue) of the high ($10^{-7}$ M$_{\odot}$) and low ($10^{-8}$\,M$_{\odot}$) mass CPDs.  The four circles indicate the present-day radial position of the Galilean satellites and their position on the ordinate is arbitrary.  The striped gray region on the left indicates an empty inner magnetic cavity and the light gray region on the right indicates the gravitationally unstable zone outside of R$_{\rm H}/3$ (where R$_{\rm H} = 0.35$\,au).}
      \label{fig:CPD_SURFACE_DENSITY}
\end{figure}

The outer radius of the CPD is set as the planetary Hill radius R$_{\rm H}$, however an exponential decline in the surface density profile is parameterized to begin at R$_{\rm H}$/3 corresponding to the outermost stable orbit set by tidal interaction and angular momentum considerations (tidal truncation radius) \citep{1998ApJ...508..707Q, 10.1111/j.1365-2966.2009.15002.x, 2011MNRAS.413.1447M}.  An empty inner magnetospheric cavity is assumed to exist as the result of magnetic interaction with the planet \citep{Takata1996,Shibaike2019,Batygin2018}.  The parameterized radial surface density profiles of the high- and low-mass CPDs can be found in Fig.\ref{fig:CPD_SURFACE_DENSITY} together with the resulting optical extinction profiles derived from the continuum radiative transfer. 

The small parameter variation grid of models exploring possible CPD mass and viscosity can be found in Table \ref{tab:cpd_grid} along with the model id's. The format of the id is ($x$-$y$) where $x$ is related to the CPD mass by \mbox{$M_{\rm CPD} =$ 10$^{-x}$ M$_{\odot}$} and $y$ to the mass infall rate (and thus viscosity) by \mbox{$\dot M_{\rm CPD} =$ 10$^{-y}$ M$_{\odot}$ yr$^{-1}$}. The list of parameters which are common between the reference CPDs can be found in Table \ref{tab:cpds}.

\subsubsubsection{CPD viscosity} \label{sec:cpdviscosity}

While the mechanism that produces angular momentum transport in accretion disks is not well understood, it is known that molecular viscosity alone is far too weak to explain observations \citep{1973A&A....24..337S,Pringle1981}.  The efficiency of angular momentum transport is parameterized by the dimensionless $\alpha$-viscosity \citep{1973A&A....24..337S} which for a circumstellar disk may have a broad distribution ranging from $\sim10^{-5}-10^{-1}$ \citep{Rafikov2017,Ansdell2018,Villenave2022}.

\begin{figure}
  \includegraphics[width=\textwidth/2]{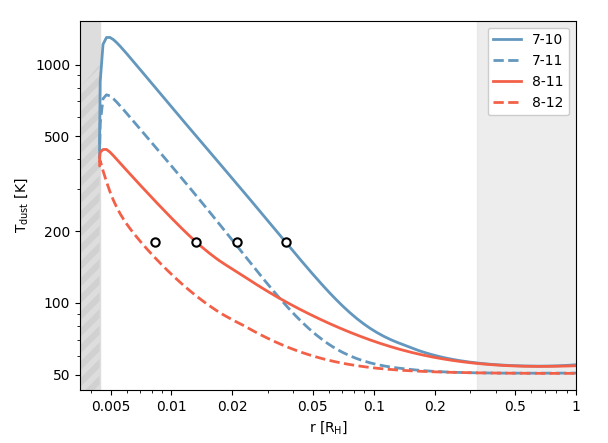}
      \caption{Radial midplane dust temperature profiles of the four reference CPDs ($d/g = 10^{-3.3}$), labelled by the model id's in Table \ref{tab:cpd_grid}. The four circles indicate only the present-day radial position of the Galilean satellites. The striped gray area on the left indicates the inner cavity.  The light gray area on the right indicates the gravitationally unstable zone outside of R$_{\rm H}/3$ (where R$_{\rm H} = 0.35$ au).}
      \label{fig:CPD_TDUST}
\end{figure}

Disk gas with a sufficiently high ionization fraction couples to the stellar or planetary magnetic field such that the  magnetorotational instability (MRI) induced turbulence may provide the source of the effective viscosity and produce \mbox{$\alpha \geq 10^{-3}$} \citep{Balbus1991,Hawley1995,Balbus2003}.  In the case of a CPD, MRI induced-turbulence may be inhibited by the short orbital periods and presence of magnetic dead-zones, limiting gas transport to a thin surface layer \citep{2014ApJ...785..101F}.   Even if the CPD were effectively inviscid,  tidal interaction with the star may promote a minimum rate of angular momentum transport through the excitation of spiral waves \citep{Rivier2012}. In our \textsc{ProDiMo} model the $\alpha$-viscosity of the disk is not explicitly specified. Instead, a mass accretion rate is specified which controls the heating rate of the disk through viscous dissipation.  

The disk mass, accretion rate, and viscosity are highly degenerate properties.  3D hydrodynamical modeling of gas delivery into the vicinity of a Jupiter mass planet at 5 au suggest \mbox{$\dot M = 10^{-9.3}$ M$_{\odot}$ yr$^{-1}$} of gas  \citep{Szulagyi2021}.  Stellar tidal perturbation may produce a minimum accretion rate \mbox{$10^{-9.7}$ M$_{\odot}$ yr$^{-1}$} \citep{Rivier2012}. In the PDS 70 system, two massive planets are observed to be accreting gas in a large dust cavity \citep{2018ApJ...863L...8W, 2018A&A...617A..44K,2019NatAs...3..749H}. \mbox{K-band} observations of PDS 70 b with the VLT are consistent with \mbox{$\dot M = 10^{-10.8} - 10^{-10.3} $ M$_{\odot}$ yr$^{-1}$} \citep{Christiaens2019} with similar values estimated for PDS 70 c \citep{Thanathibodee2019}.  HST UV and H$\alpha$ imaging of the protoplanet PDS 70 b suggest \mbox{$\dot M = 10^{-10.9} - 10^{-10.8}$ M$_{\odot}$ yr$^{-1}$} \citep{Zhou2021}. Based on these observational and theoretical constraints we adopt \mbox{$\dot M = 10^{-10}$ M$_{\odot}$ yr$^{-1}$} (with a heating rate corresponding to \mbox{$\alpha \approx 10^{-2.7}$}) and \mbox{$\dot M = 10^{-11}$ M$_{\odot}$ yr$^{-1}$} \mbox{($\alpha \approx 10^{-3.6}$)} for the high-mass CPD, representing viscous timescales of $10^3$ and $10^4$ years, respectively, over which the majority of the CPD mass is replaced by freshly accreted material (see Eq.(\ref{eq:tvisc})).  For the low-mass CPD we adopt \mbox{$\dot M = 10^{-11}$ M$_{\odot}$ yr$^{-1}$} and \mbox{$\dot M = 10^{-12}$ M$_{\odot}$ yr$^{-1}$} to represent the same $\alpha$-viscosities and $t_{\rm visc}$.

The viscous heating is determined according to the method of \citet{1998ApJ...500..411D}. The half-column heating rate is 

\begin{equation} \label{eq:visc_heating}
    F_{\rm vis} (r) = \frac{3 G M_{\rm p} \dot{M}}{8 \pi r^3} (1-\sqrt{R_{\rm p} /r})   \hspace{0.1cm} \textrm{erg cm}^{-2}  \textrm{s}^{-1},
\end{equation} 

\noindent
where $G$ is the gravitational constant, $M_{\rm p}$ is the planet mass, $r$ is the distance to the planet, and $R_{\rm p}$ is the planetary radius. We convert the surface-heating to a heating rate per volume as

\begin{equation}
    \Gamma_{\rm vis}(r,z) =  F_{\rm vis} (r) \frac{\rho_{\rm d}(r,z)}{\int \rho_{\rm d}(r,z') dz'} \textrm{erg cm}^{-3}  \textrm{s}^{-1},
\end{equation}
where $\rho_{\rm d}$ is the mass density of the dust at radius $r$ and height $z$.  The heating is applied directly to the dust. The resulting midplane dust temperature profile of the CPDs can be found in Fig.\ref{fig:CPD_TDUST}. The temperature profiles have been calculated using a new diffusion-approximation radiative transfer solver which is described in appendix \ref{appendix:rtdiffusionsolver}.

\subsubsubsection{Sources of CPD external irradiation}

The rate of photodesorption plays an important role in determining the ice abundance in the outer optically-thin region of the CPDs.  Potential sources of radiation include the planet,  the star, and nearby massive cluster stars. For the planet we assume the runaway gas accretion phase has ended and that the luminosity has correspondingly declined to $10^{-5}$\,L$_{\odot}$ with a surface temperature of 1000\,K where it remains relatively constant for 10 Myr \citep{Hubickyj2005, 2007ApJ...655..541M, 2012ApJ...745..174S}.  

We parameterize the isotropic background radiation intensity with the dimensionless parameter $\chi$.  The background intensity is then the sum of a diluted 20000\,K blackbody field and the cosmic microwave background,
\begin{equation} \label{eq:chi}
    I_{\nu}^{bg} = \chi \cdot 1.71 \cdot W_{\rm dil} B_{\nu}(20000 K) + B_{\nu}(2.7 K),
\end{equation}
\noindent
where the dilution factor \mbox{$W_{\rm dil}$ = $9.85357\times10^{-17}$} such that a value of $\chi = 1$ corresponds to the quiescent interstellar background, or "unit Draine field" \citep{Draine1996, Roellig2007, Woitke2009a}.  Irradiation by the young sun provides a minimum $\chi\,\sim\,3000$ at 5\,au (see Fig.\ref{fig:mmsn-gap-conditions}) despite partial shadowing by an inner disk \citep{Oberg2020}.  We adopted the same value for the strength of the FUV irradiation in the midplane at Jupiter's location although it is contingent on our assumptions regarding the stellar UV luminosity and geometry of the inner circumstellar disk.  Independently of these factors, \citet{Oberg2020} find that interstellar radiation results in a mean $\chi$ of O(3) in the gap, as the Sun is believed to have formed in a relatively dense stellar cluster \citep{Adams2010,PZ2019} containing massive OB stars  \citep{2013A&A...549A..82P}.  External irradiation heats dust and gas on the surface and in the outer regions of the optically-thin CPD midplane. 3D dust radiative transfer modelling of gap-embedded CPDs suggests scattered stellar radiation can be the dominant heating source in the outer regions of a CPD \citep{Portilla-Revelo2022}.  We assumed that the outer edge of the CPD is in thermal equilibrium with the surroundings and set a CPD background temperature lower limit of 50\,K equal to that of dust in the circumstellar disk gap (see Fig. \ref{fig:mmsn-gap-conditions}).

\subsubsubsection{CPD dust mass and grain size population}

The dust-to-gas ratio is a key parameter that regulates the eventual ratio of ice to rock in the CPD.  Planet-induced pressure bumps at the gap edges of the circumstellar disk may filter out dust particles above $\sim$100 \textmu m in size \citep{2006MNRAS.373.1619R,Zhu2012}.  For a dust grain size population power law exponent of -3.5, minimum grain size a$_{\rm min}$ \mbox{0.05 \textmu m}, maximum grain size a$_{\rm max}$ \mbox{3000 \textmu m}, a filtering of grains larger than 100 \textmu m  would be reduced the mass of dust entering the gap by a factor $\sim30$.  As an opposing view \citet{Szulagyi2021} find that a planet within the gap can stir dust at the circumstellar disk midplane and produce a very high rate of dust delivery to the CPD, such that the dust-to-gas ratio can even be enhanced in the CPD to $\sim\,0.1$ for a Jupiter mass planet at 5\,au. 

The dust population within the CPD may also rapidly evolve.  It is possible that mm-sized grains are quickly lost to radial drift in $10^2-10^3$ yr \citep{Zhu2018}.  Similarly \citet{Rab2019} used the dust evolution code \textit{two-pop-py} \citep{Birnstiel2012} to show that for an isolated CPD onto which new material does not accrete, an initial dust-to-gas ratio of $10^{-2}$ can in only $10^4$\,yr be reduced to $< 10^{-3}$ and in $10^5$\,yr to $<\,10^{-4}$.   However, larger dust grains may become trapped in CPDs  \citep{2018ApJ...866..142D, Batygin2020}.   Additionally, we expected that in an embedded, actively-accreting CPD the dust will continually be replenished and that a higher steady-state dust-to-gas ratio will be achieved.  Given these considerations we tested various possible dust to gas ratios ranging from \mbox{$10^{-4}-10^{-2}$} in Appendix \ref{appendix:dusttogas}.

\begin{table}
    \caption{Parameters common to the CPD models. Parameters which are not common to the CPD models are listed in Table \ref{tab:cpd_grid}. Where not specified the CPD parameters are identical to the circumstellar disk parameters listed in Table \ref{tab:ppds}.  }

    \centering
    \renewcommand{\arraystretch}{1.1}%

   \begin{tabular}{llll}
    \hline \hline
        Parameter               & Symbol              & Value         & Unit          \\ \hline  
        Planet Mass             & $M_{\rm p}$         & 1.0           &  M$_{\rm J}$  \\
        Planetary Luminosity    & $L_{\rm p}$         & $10^{-5}$     &  L$_{\odot}$  \\
        
        Effective Temperature   & $T_{\rm eff,p}$     & 1000          & K           \\
        UV Luminosity           & $L_{\rm UV,p}$      & 0.01          & L$_{\rm p}$ \\
        Background UV Field     & $\chi$              & $3\times10^3$ & -           \\ 
        Background Temperature  & $T_{\rm back}$      & 50            & K           \\

        \hline

      %  Disk Mass                 & M$_{\rm CPD}$        & 10$^{-7}, 10^{-8}$ & M$_{\odot}$ \\
        Disk Inner Radius         & $R_{\rm in,CPD} $    & 0.0015 & au \\
        Taper Radius              & $R_{\rm tap,CPD} $   & 0.12   & au \\ 
        Disk Outer Radius         & $R_{\rm out,CPD} $   & 0.35   & au \\
        Column Density Power Ind. & $\epsilon_{\rm CPD}$ & 1.0    & -  \\
        Flaring Index             & $\beta_{\rm CPD}$    & 1.15   & -  \\
        Reference Scale Height    & $H_{\rm 0.1 au}$     & 0.01   & au \\
        
        \hline
        
        Maximum dust size     & $a_{\rm max,CPD}$  & 3000        & $\mu$m   \\
        Dust-to-Gas Ratio     & $d/g$          & $10^{-3.3}$ & -        \\

     %   \hline
        
      %  Accretion rate     &  $\dot M$ & $10^{-13} - 10^{-9} M_{\odot}$ yr$^{-1}$ \\ \hline

    \end{tabular}

    \label{tab:cpds}
    
\end{table}

\begin{table}
    \caption{Variation of parameters for the circumplanetary disk model grid. Model id's are of the format (A-B) where the CPD mass is $10^{\rm -A}$ M$_{\odot}$ and where the accretion rate onto the CPD is $10^{\rm -B}$ M$_{\odot}$ yr$^{-1}$.}

    \centering
    \renewcommand{\arraystretch}{1.1}%

   \begin{tabular}{lllll}
    \hline \hline
    \vspace{-2.0ex} \\
    model id & $M$ [M$_{\odot}$] & $\dot{M}$ [M$_{\odot}$yr$^{-1}$] & $t_{\rm visc}$ [yr] & $\alpha$ \\ \hline
    \vspace{-2ex} \\
        (7-10) & $10^{\textbf{-7}}$ & $10^{\textbf{-10}}$ & $10^3$ & $10^{-2.7}$ \\
    \vspace{0.2ex}
        (7-11) & $10^{\textbf{-7}}$ & $10^{\textbf{-11}}$ & $10^4$  & $10^{-3.6}$ \\
    \vspace{0.2ex}
        (8-11) & $10^{\textbf{-8}}$ & $10^{\textbf{-11}}$ & $10^3$ & $10^{-2.7}$ \\
    \vspace{0.2ex}
        (8-12) & $10^{\textbf{-8}}$ & $10^{\textbf{-12}}$ & $10^4$ & $10^{-3.6}$ \\
    \end{tabular}

    \label{tab:cpd_grid}
    
\end{table}

\subsection{Dust grain drift within the CPD} \label{sec:methods:drift}

In \textsc{ProDiMo} the chemistry is solved for each grid cell without accounting for radial dust or gas transport.  Instead we used properties of the \textsc{ProDiMo} model output to inform the radial dust drift calculations in a post-processing step to compare timescales of chemical evolution vs. dust drift. Disk gas which is pressure-supported orbits at sub-Keplerian velocities such that larger grains feel a headwind and rapidly drift inwards \citep{Weidenschilling1977}.  In a circumstellar disk the degree of sub-Keplerianity can be $<1\%$ while in a CPD it could be as high as 20-80$\%$ due to significant gas pressure support \citep{Armitage2007,2018ApJ...866..142D}.   We considered whether the timescale of grain drift can compete with the processes that shape grain composition  In the Epstein regime the radial grain drift velocity $v_{r,d}$ can be approximated as

\begin{equation}
    v_{r,d} = \frac{ v_{r,g} T_s^{-1} - \eta v_K}{T_s + T_s^{-1}} , 
\end{equation}
\noindent
where $v_{r,g}$ is the radial velocity of the gas, $T_s$ is the dimensionless stopping time of a grain, 

\begin{equation}
    T_s = t_s \bigg( \frac{v_k}{r} \bigg) , 
\end{equation}
\noindent
where $t_{\rm s}$ is the stopping time, $v_{\rm k}$ is the Keplerian orbital velocity, and $r$ is the radius in the CPD.  The stopping time is 

\begin{equation}
    t_s = \bigg( \frac{\rho_{\rm grain}}{\rho_{\rm gas}} \bigg) \bigg( \frac{a}{v_{\rm th}} \bigg) , 
\end{equation}
\noindent
where $a$ is the grain size, $\rho_{gas}$ is the gas density, $\rho_{\rm grain}$ is the material density of the dust grains, and $v_{\rm th}$ is the thermal velocity of the gas: $c_{\rm s}$ (8/$\pi)^{0.5}$ where $c_s$ is the speed of sound.  The parameter $\eta$ is 

\begin{equation}
    \eta = n \bigg( \frac{c_s^2}{v_k^2} \bigg) , 
\end{equation}
\noindent
and $n$ is the power law exponent of the radial pressure gradient \citep{Armitage2010}. We extract the gas density $\rho_{\rm gas}$, sound speed $c_{\rm s}$, and pressure gradient from our disk models to consistently determine the grain drift velocities for a grain material density \mbox{$\rho_{\rm grain} = 3$ g cm$^{-3}$}. The Epstein regime is valid where the dust particle size $a \leq 9 \lambda / 4$ where $\lambda$ is the mean free path of the gas particles.  In the inner CPD the gas density is sufficiently high that this condition can be violated.  For the high-mass CPD this occurs inside of the orbit of Callisto and for the low-mass CPD inside of the orbit of Europa for grains less than 1 cm in size, in which case we transition to the Stokes regime and calculate the drift velocities accordingly \citep{Weidenschilling1977}.

We adopted several simplifying assumptions regarding the radial velocity structure of the gas.  The centrifugal radius $r_{\rm c}$ of the CPD is the orbital radius at which the specific angular momentum is equal to the average of the infalling material and where solid material is accreted onto the CPD \citep{Canup2002,2010ApJ...714.1052S}. In the case of a Jupiter-mass planet this lies near the orbit of Callisto. Rather than accreting at precisely this radius infalling matter have some intrinsic spread in angular momentum \citep{2008ApJ...685.1220M}.  We adopted the prescription of \citet{2011AJ....142..168M} that material will accrete between 16-28 R$_{\rm J}$. The gas falling onto the CPD at $r_{\rm c}$ will viscously spread radially away from where it is accreted \citep{Pringle1981}.  Hence interior to $r_{\rm c}$ gas flows towards the planet and exterior to $r_{\rm c}$ it will flow outwards. Recently it has been proposed that to effectively trap dust and allow for planetesimal growth within the CPD, gas may indeed need to flow predominantly away from the planet and thus form a decretion disk \citep{Batygin2020}.  

In our high viscosity models, a parcel of gas that accretes onto the CPD near $r_{\rm c}$ and flows outwards must travel $\sim 0.3$ au within $10^3$ yr to be consistent with $t_{\rm visc}$ and thus has a mean outwards radial velocity on the order of \mbox{1 m s$^{-1}$}.  For our low viscosity case \mbox{$v_{\rm r, gas} \sim 0.1 $ m s$^{-1}$}.

\section{Results} \label{sec:results}

In our reference model grid (Table \ref{tab:cpd_grid}) we considered the case of a higher mass, optically thick CPD \mbox{($10^{-7}$\,M$_{\odot}$)}, and lower mass, optically thin CPD ($10^{-8}$\,M$_{\odot}$) both with $d/g = 10^{-3.3}$.  For each unique mass we considered viscous timescales of $10^{3}$ and $10^{4}$\,yr.  For each of the four resulting CPDs we tested two initial compositions: of chemical inheritance and reset, for a total of eight models.

\subsection{Timescales of ice formation} \label{sec:results_timescales}

In the following section we discuss by which pathways and at what rate water (ice) formation occurs in a chemically reset CPD. Thereafter we contrast these results with that of the CPDs which inherit an initial chemical composition from the circumstellar disk.

\subsubsection{Reset}

\begin{figure}[ht!]
  \includegraphics[width=\textwidth/2]{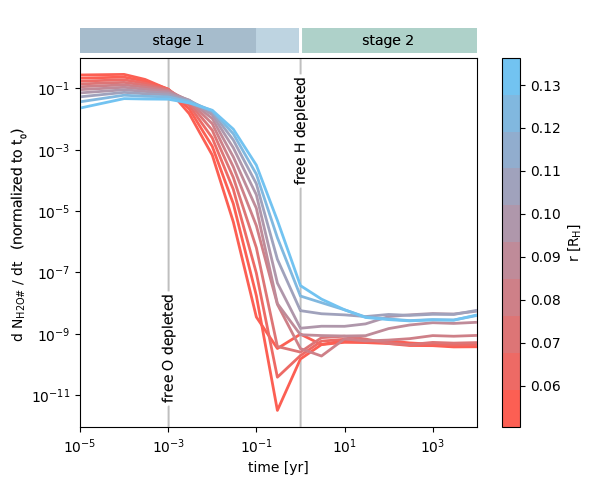}
      \caption{Rate of water ice formation as a function of time at different radii beyond the snowline in the high-mass, high-viscosity CPD (7-10), to illustrate the two distinct stages of water ice formation.  All values are normalized to the maximum formation rate at 10$^{-5}$\,yr, r = 0.08\,R$_{\rm H}$.  The times where the decline in the abundance of free O and free H end are indicated by the gray vertical lines. The starting time $t_0$ is defined as when the infalling gas is shock-heated to an atomic and ionized state. }
      \label{fig:dnice_dt}
\end{figure}

In our ``reset" scenario the species initially present are exclusively atomic (with the exception of PAHs) and singly or doubly ionized.  All hydrogen is initially present in the form of H$^+$.  The reset case is characterized by the formation of ice where it is stable against desorption.  Ice formation is suppressed in the innermost region of the CPD due to dust temperatures in excess of 160-180\,K. In the outer region of the CPD ice formation is suppressed by the low optical depths and correspondingly high intensity of background radiation which causes ices to photodesorb.  Between these two boundaries ices begin to form

The freeze-out occurs step-wise, with two characteristic timescales on which ice formation proceeds. The rate of water ice formation at different disk radii as a function of time is shown in Fig. \ref{fig:dnice_dt}.  Within \mbox{$0.01-1$\,yr} the first step is completed. This initial, rapid water formation occurs primarily via a path that begins with a three-body recombination reaction important at temperatures below 2800\,K \citep{Hidaka1982,Tsang1986},

\vspace*{-1mm}

\begin{equation} \label{eq:water1a}
    \rm H + O + M \rightarrow OH + M,
\end{equation}

\noindent
where the third body M = H, H$_2$, or He, and to a lesser extent by \mbox{H + O $\rightarrow$ OH + photon}. The water then forms by radiative association of the OH with free hydrogen;

\vspace*{-1mm}

\begin{equation} \label{eq:water1c}
    \rm OH + H \rightarrow H_2O + photon,
\end{equation}

\noindent
after which it adsorbs to a grain.  Water formation via the reactions \ref{eq:water1a} and \ref{eq:water1c} remains proportional to the declining abundance of free H and ends when it is depleted around $10^{-1}$\,yr.  Typically half of the maximum possible amount of water ice that could form is produced during this stage.  

The first stage of water ice formation proceeds inside-out as a result of the strong density-dependence of the collider reaction. The second stage proceeds outside-in due to the pathway's dependence on intermediate species produced by photo-reactions. This can be seen in Fig. \ref{fig:dnice_dt} where the inner disk formation rate (red lines) is initially higher while later the outer disk rate (blue lines) is relatively higher.  At this lower rate of formation the total mass of water ice doubles at the midplane within $\sim1$\,Myr, approximately half of which forms by

\vspace*{-1mm}

\begin{equation}
    \rm NH_2 + NO \rightarrow N_2 + H_2O,
\end{equation}

\noindent
near the snowline. The second stage of ice formation also involves the freezing-out of NH$_3$ and other more volatiles species.  We find that the relatively unstable NH$_2$ exists in abundance at such high densities ($n_{\rm H} \sim 10^{15}$\,cm$^{-3}$) due to the three-body reaction \mbox{N + H$_2$ + M $\rightarrow$ NH$_2$ + M}.  Although the reaction rate has been experimentally determined \mbox{$k_0$ = 10$^{-26}$ cm$^6$ s$^{-1}$} \citep{Avramenko1966}, it has been noted as being in need of revisiting due to its importance in water formation via \mbox{NH$_2$ + O $\rightarrow$ NH + OH} \citep{Kamp2017}.  We have adopted a rate more typical of collider reactions \mbox{(10$^{-30}$ cm$^6$ s$^{-1}$)}, which still produces enough NH$_2$ for this path to play an important role.  The other half of water ice is formed via the more ``classical" route

\vspace*{-1mm}

\begin{equation}
    \rm H_3O^+ + e^- \rightarrow H_2O + H,
\end{equation}

\noindent
and in the outermost part of the disk where water ice is stable via 

\begin{equation}
    \rm H_3O^+ + H_2CO \rightarrow H_3CO^+ + H_2O,
\end{equation}

\noindent
In the high-mass CPDs water ice formation is typically still ongoing by the end of the viscous timescale, and so the midplane ice abundance is not able to converge to the level seen in steady-state within the viscous timescale.

\subsubsection{Inheritance}

The evolution of the inheritance case is characterized by the desorption of ices in regions where they are not stable against thermal or photo-desorption.  Where water ice is stable very little additional water formation occurs within the viscous timescale.

Icy grains interior to the snowline sublimate typically within $10^{-5}$\,yr, and a ``snowline" (water iceline) is clearly established. In some cases the snowline can take longer to stabilize, shifting outwards over time, and sometimes continues to evolve radially for up to $10^5$ yr.  This is notable in the (8-11) CPD where the snowline moves from 0.01\,R$_{\rm H}$ at $10^{-5}$\,yr to 0.03\,R$_{\rm H}$ after $10^4$\,yr. In practice this implies that there exists a radial span in which the composition of radially drifting icy grains does not immediately begin to reflect the ambient conditions.  This is discussed in Sect.  \ref{sec:discussion_drift}.

In the outer optically thin region of the CPDs, ices are also lost to photodesorption although the process is slower than the thermal desorption occurring in the inner disk.  Desorption in this area is typically complete within $10^3-10^5$\,yr and has not always reached steady-state by the end of the viscous timescale.

\begin{figure*}[ht!]

\begin{subfigure}{\textwidth/2}
    \centering
  \includegraphics[width=\textwidth]{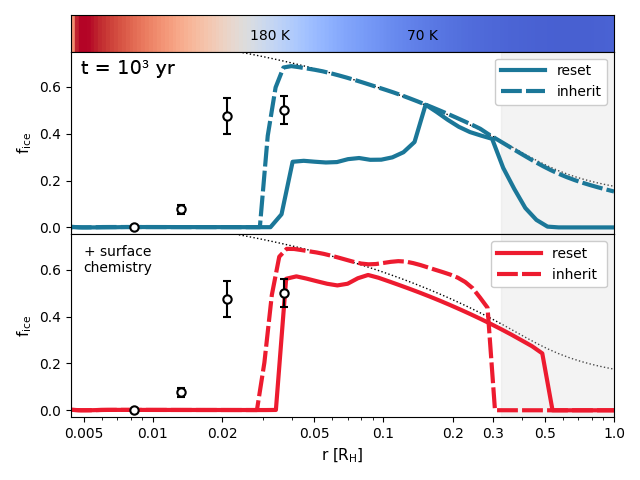}
      \caption{High-Mass High-Viscosity (7-10)}

\end{subfigure}%
\begin{subfigure}{\textwidth/2}
    \centering
  \includegraphics[width=\textwidth]{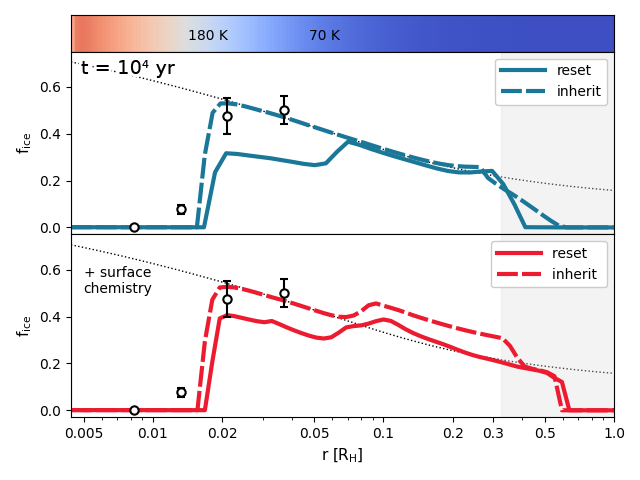}
      \caption{High-Mass Low-Viscosity (7-11)}

\end{subfigure}

\begin{subfigure}{\textwidth/2}
    \centering
  \includegraphics[width=\textwidth]{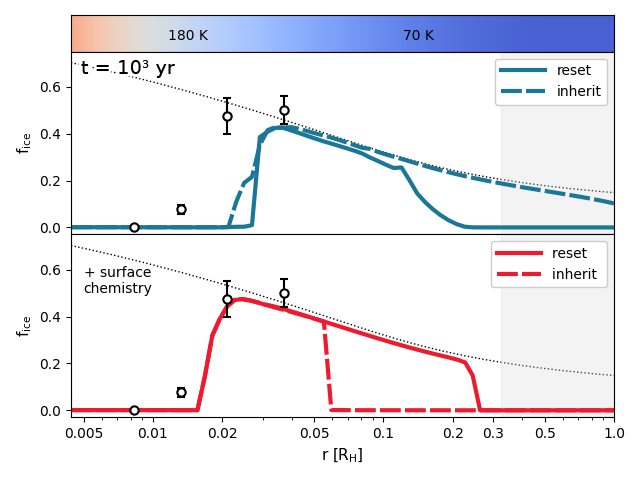}
      \caption{Low-Mass High-Viscosity (8-11)}

\end{subfigure}%
\begin{subfigure}{\textwidth/2}
    \centering
  \includegraphics[width=\textwidth]{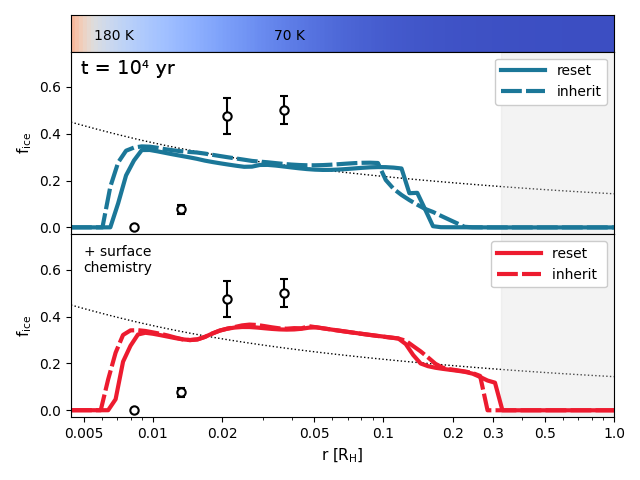}
      \caption{Low-Mass Low-Viscosity (8-12)}

\end{subfigure}

\caption{Midplane ice fraction at $t = t_{\rm visc}$  for the CPDs with \mbox{$d/g = 10^{-3.3}$}.  The four circles indicate the present day radial location of the Galilean satellites and the uncertainty bars represent their possible range of ice fractions. The thin dotted line indicates the initial radial ice fraction in the inheritance cases.  The filled gray region on the right indicates the gravitationally unstable zone outside of R$_{\rm H}/3$.}
\label{fig:fice_all4}
\end{figure*}

\subsection{The midplane ice mass fraction}

While water ice can form efficiently in chemically reset CPDs, the final $f_{\rm ice}$ of the solids depends strongly on the total dust mass in the midplane.   We explore the role of the global $d/g$ ratio in the reset and inheritance cases in Appendix \ref{appendix:dusttogas}, Fig. \ref{fig:fice_dustgas_001}.  A canonical dust-to-gas ratio of $10^{-2}$ produces at most grains with an ice mass fraction of $<0.1$ and is nowhere consistent with the composition of the icy Galilean satellites.   In contrast a dust-to-gas ratio of $10^{-3.3}$ results in solids with $f_{\rm ice}$ both above and below the maximum Galilean $f_{\rm ice}$ for the high and low-mass CPDs, respectively.   Hereafter the global dust-to-gas ratio of $10^{-3.3}$ is considered in discussions of the four reference CPDs.  

We show the state of the radial ice mass fraction for the CPDs with $d/g = 10^{-3.3}$ on their respective viscous timescales to allow for direct comparison between the inheritance and reset cases in Fig.\ref{fig:fice_all4}.  For completeness we include the results where grain-surface chemical reactions are utilized.

The midplane $f_{\rm ice}$ is also dependent on the degree of dust sedimentation (settling). A general feature of the $f_{\rm ice}$ profiles is an inner maximal peak at the snowline followed by a decline towards the outer edge of the disk.  As dust settling is more efficient at larger radii,  $f_{\rm ice}$ reduces accordingly with increasing radius.  Settling of dust to the midplane is counteracted by stochastic advection by turbulent eddies in the gas. We assume that the value of turbulent-$\alpha$ used to determine the degree of settling is equal to the global viscous $\alpha$ used to determine the heating rate by viscous dissipation. In the low viscosity CPDs (7-11) and (8-12) dust settling towards the midplane is thus proportionally enhanced.  In the low-viscosity cases the dust density is enhanced in the midplane minimally by a factor $\sim 3$ over the global \textit{d/g}, increasing to a factor $\sim$20 at R$_{\rm H}/3$.  As the degree of settling is also dependent on the adopted surface density power law exponent $\epsilon$ we explore the impact of deviation from the assumed $\epsilon=1$ in Appendix \ref{appendix:surfacedensityslope}. Given that we have no reason to believe this value will depart significantly from the range 1.0-1.3, the resulting $f_{\rm ice}$ in the inner disk should differ from our reference result by no more than 25-30$\%$ interior to the ammonia iceline at $\sim$70\,K.

In general the high-mass chemically reset CPDs (7-10) and (7-11) are not able to converge entirely towards a steady-state ice abundance in either the ``fast" ($10^3$\,yr) or ``slow" ($10^4$\,yr) viscous timescales as gas-phase CO is more stable and contains a relatively larger fraction of the total oxygen budget for longer.  As a result chemically reset CPDs contain on average less water ice than those which inherit their ices from the circumstellar disk. In contrast, the low-mass chemically reset CPDs (8-11) and (8-12) are able to converge towards the ice abundances seen in the inheritance cases within 100 yr.

\begin{figure}
  \includegraphics[width=\textwidth/2]{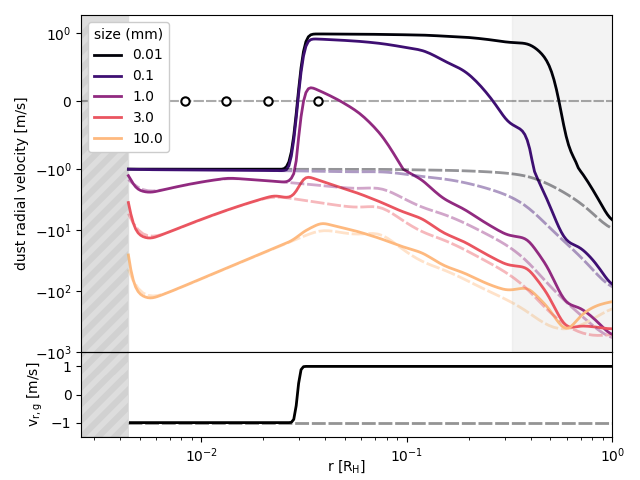}
      \caption{Radial drift velocity of particles (\textit{upper panel}) and radial velocity of the gas (\textit{lower panel}) in the high-mass, high-viscosity CPD (7-10).  The position of the Galilean satellites on the ordinate is arbitrary.  The gray region on the right indicates the tidally unstable region beyond R$_{\rm H}/3$. The striped region on the left indicates the inner cavity. Solid lines indicate the case of a decretion disk where gas falls onto the CPD at the centrifugal radius (0.03\,R$_{\rm H}$). Dashed lines indicate the grain drift and gas radial velocity in the case of a pure accretion disk.}
      \label{fig:drift_velocity}
\end{figure}

\subsubsection*{The role of surface chemistry} \label{sec:results_surface}

The duration of the initial stage in which water formation is rapid is dependent on the availability of atomic H.  When  H$_2$ formation is complete this stage ends. The formation of H$_2$ is treated differently with the inclusion of grain surface chemistry.  When the chemistry is limited to gas-phase reactions and direct adsorption/desorption only, H$_2$ formation proceeds via the analytic rate determined by \citet{Cazaux2002}.  When surface chemistry is included H$_2$ formation is instead modeled explicitly via reactions involving hydrogenated PAHs (H-PAH) \citep{Boschman2012,Thi2020H2}.

A chemical reset poses a scenario in which H$_2$ and H$_2$O formation occur simultaneously.  The analytic rate of \citet{Cazaux2002} presupposes that chemisorbed H plays a role in H$_2$ formation on silicate or carbonaceous surfaces, in which H goes through an intermediate stage of being chemically, rather than physically, bound to the grain surface.  We find that prior to atomic H depletion several (>3) monolayers of H$_2$O have formed on the average sized grain. The formation of H$_2$ via chemisorbed H should thus be suppressed in these regions as the water layers prevent H atoms from chemisorbing to the grain surface \citep{WAKELAM2017B}.  In the absence of chemisorbed H, H$_2$ formation on dust is strongly reduced and H$_2$ formation proceeds primarily via H-PAH. The H$_2$ formation rate under these conditions is less efficient than the analytic rate from \citep{Cazaux2002} near the inner edge of the snowline at 150-170\,K. The relatively slower formation of H$_2$ and the resulting prolonged availability of atomic H results in a $\sim 30-100\%$ increase in water ice abundance interior to the NH$_3$ iceline prior to $t_{\rm visc}$ and hence narrows the gap between the inheritance and reset cases in this region. Water ice formation in the inner disk is further enhanced by the inclusion of O$_2$H in the  surface chemistry network.  Gas-phase three-body reactions with O$_2$ produce O$_2$H which in turn lead to early OH formation.  The gas-phase O$_2$ reservoir is thus depleted and efficiently converted via OH into H$_2$O via the three-body reaction

\begin{equation}
    \rm O_2 + H + M \rightarrow O_2H + M,
\end{equation}
\noindent
with rates adopted from UMIST2006 \citet{Atkinson2004,Woodall2007}, which is highly efficient at the densities in the CPD, and thereafter

\begin{equation}
    \rm O_2H + H \rightarrow OH + OH,
\end{equation}

\begin{equation}
    \rm OH + H \rightarrow H_2O + photon,
\end{equation}

In the outer disk the ice mass fraction can be enhanced relative to the gas-phase chemical network as the freezing-out of more volatile species is facilitated by grain surface reactions.  CO$_2$ ice is readily formed on the surface via \mbox{OH + CO $\rightarrow$ CO$_2$ + H} \citep{Oba2010,Liu2015} for which we adopt an  effective barrier of 150\,K \citep{Fulle1996,Ruaud2016}.   The formation of carbon bearing ices begins to significantly influence the $f_{\rm ice}$ of the chemically reset CPDs only after 10$^3$\,yr and hence the effect on the high-viscosity CPDs with $t_{\rm visc} = 10^3$\,yr is less pronounced.  The formation and composition of these ice impurities will be discussed in detail in an accompanying work (Oberg et al. in prep).

\begin{figure*}[ht!]

\begin{subfigure}{\textwidth/2}
    \centering
  \includegraphics[width=\textwidth]{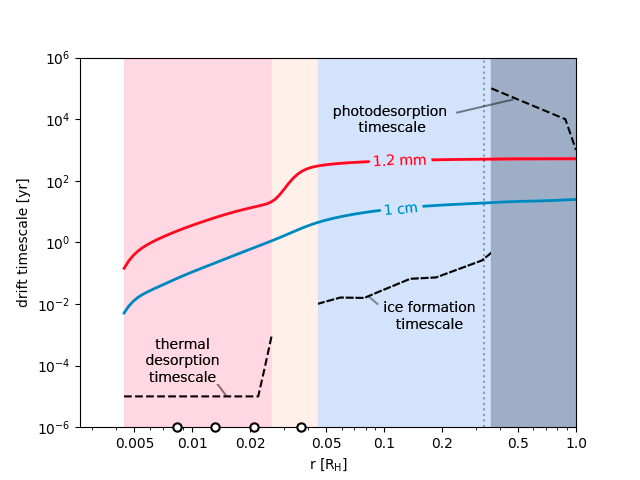}
      \caption{High-Mass High-Viscosity (7-10)}

\end{subfigure}%
\begin{subfigure}{\textwidth/2}
    \centering
  \includegraphics[width=\textwidth]{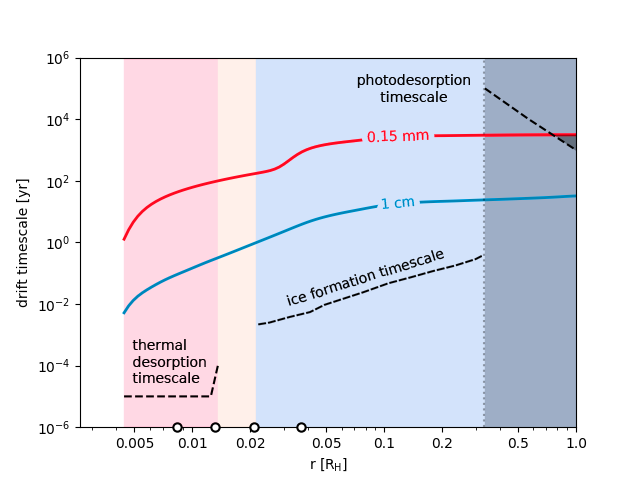}
      \caption{High-Mass Low-Viscosity (7-11)}

\end{subfigure}

\begin{subfigure}{\textwidth/2}
    \centering
  \includegraphics[width=\textwidth]{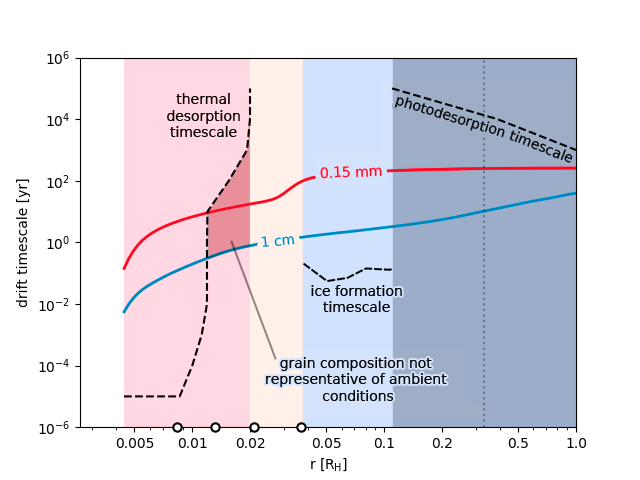}
      \caption{Low-Mass High-Viscosity (8-11)}

\end{subfigure}%
\begin{subfigure}{\textwidth/2}
    \centering
  \includegraphics[width=\textwidth]{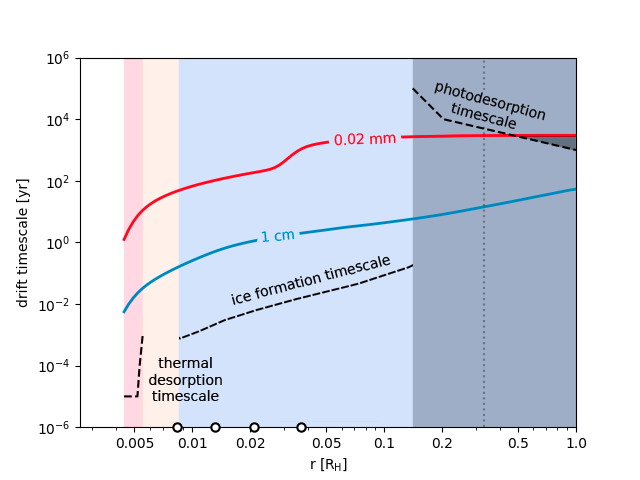}
      \caption{Low-Mass Low-Viscosity (8-12)}

\end{subfigure}

\caption{Inwards drift timescale for dust grains (colored lines) relative to the timescales of desorption and ice formation (black dashed lines) in the reference CPDs with \mbox{$d/g = 10^{-3.3}$}. From left to right the colored regions correspond to areas where all ices are eventually thermally desorbed (pink), the iceline (light yellow), where grains become icier (light blue) and where all ices are eventually photodesorbed (dark gray).  The red line corresponds to the minimum grain size which is not trapped in the CPD (if there is a sign change in the gas radial velocity) and the blue line to the maximum for which the timescales of desorption and ice formation have been verified.  The vertical dotted line indicates the approximate outer region of gravitational stability in the CPD.}
\label{fig:drift_time}
\end{figure*}

\subsection{Grain drift vs. adsorption and desorption} \label{sec:grain_drift}

We calculated the velocity of radially drifting grains in the four reference CPDs and showcase the results for the high-mass high-viscosity CPD (7-10) in Fig. \ref{fig:drift_velocity}. We solved for the total time it takes grains of several sizes to reach the inner disk edge.  The resulting timescale of grain drift can be seen in Fig. \ref{fig:drift_time} which shows the time for a grain deposited at radius $r$ to reach the CPD inner edge. Grains which become trapped are not included in Fig. \ref{fig:drift_time}, as they do not reach the disk inner edge.  The regions where thermal desorption, ice formation, and photodesorption predominantly shape grain mantle composition, as well as their respective timescales, are indicated in the figure.  These are the timescales with which grain drift competes. 

The regions have been defined as follows:  the ``thermal desorption region" is found interior to the snowline where inherited, initially icy grains lose their icy mantles within the viscous timescale.  The ``snow border'' is the region where the reset and inheritance cases are not able to converge towards a common snowline within $10^6$ yr.  Grains here are able to retain their icy mantles but no significant additional ice adsorption occurs. The ``ice formation region" is where water begins to adsorb to grains after the CPD has been chemically reset. The ``photodesorption region" is the optically thin region exterior to the snowline where inherited grains eventually lose their icy mantles within $\sim10^6$ yr at most.  

We focus on grains of size $<10$ mm as the adsorption and desorption timescales derived in Sect.  \ref{sec:results_timescales} have only been derived with the thermochemical disk model for grains up to this size. In most cases the grain drift timescale $t_{\rm drift}$ is longer than the timescales of thermal desorption and the timescale of rapid ice formation.  The composition of grains will thus correspond to the $f_{\rm ice}$ profiles derived in Sect.  \ref{sec:results} except in the case of the low-mass, high-viscosity CPD (8-11).  In this CPD icy grains can cross into the "thermal desorption region" but only begin desorbing once they approach the position of Europa.

\subsubsection*{Dust traps}

It is clear from Fig.\ref{fig:drift_velocity} that in a decreting CPD dust traps are present.  Grains deposited near the centrifugal radius drift outwards with the gas until the force of radial advection is balanced by the loss of orbital energy from drag against gas orbiting at sub-Keplerian velocity.  Trapped grains thus become radially segregated by size, with smaller grains drifting to the outer edge of the trap and larger grains remaining near the inner edge.  In the high-mass high-viscosity CPD (7-10) grains 0.1-1\,mm in size become trapped near 0.1-0.2\,R$_{\rm H}$.  Grains smaller than 0.05\,mm advect with the gas and are able to reach the outermost stable orbit at R$_{\rm H}/3$ where they would be eventually lost to e.g. tidal stripping. 

In general the dust traps are spatially coincident with the ice formation region and in the lower-mass CPDs also partially with the photodesorption region. Hence continued ice deposition on trapped grains could facilitate grain growth.  This phenomena is discussed further in appendix Sect. \ref{appendix:trapped-grains}.

\section{Discussion} \label{sec:discussion}

We set out to explore the process of ice formation in CPDs with relatively short viscous timescales to constrain their physical, chemical, and dynamical properties.  We find that even if infalling gas and ice is fully atomized, re-freezing proceeds quickly in the CPD and solids reach an $f_{\rm ice}$ $\sim0.5$ by $t_{\rm visc}$ for an appropriate midplane dust-to-gas ratio. The midplane ice abundance at $t_{\rm visc}$ is generally insensitive to the initial chemical conditions.  Only in the inner disk ($r<0.05-0.1$\,R$_{\rm H}$) of the high-mass CPDs is ice formation too slow for the reset and inheritance cases to converge. With our standard chemical network the efficiency of water production in this region is closely tied to the availability of atomic H and thus to the H$_2$ formation rate, which is not well constrained in these conditions.  The three-body reaction \mbox{H + O + M $\rightarrow$ OH + M} is also critical to the process.  For this reaction we have adopted the rate coefficients listed in UMIST2006 \citep{Woodall2007}, the value of which originates from a recommendation of \citet{Tsang1986} who noted that literature values represented only rough estimates and suggested a factor 10 uncertainty \citep{Baulch1972,DAY1973}.  More recent estimates of this rate at high temperatures  (\mbox{> 3000 K}) suggest this value is accurate to within $\sim$40$\%$ \citep{Javoy2003}, but modern low temperature measurements remain desirable.  In our expanded grain-surface chemical network, gas-phase O$_2$H plays an important role in accelerating OH formation in the inner disk, diverting atomic oxygen into water rather than O$_2$. 

\subsection{Constraints on CPD properties}

For reasonable assumptions regarding the properties of the CPD the snowline can fall very near the present-day position of Europa. The CPD with mass \mbox{$10^{-7}$ M$_{\odot}$} \mbox{($10^{-4}$ M$_{\rm J}$)},  \mbox{$\alpha = 10^{-3.6}$} and \mbox{$d/g = 10^{-3.3}$} matches best the compositional gradient of the Galilean moons at their present-day orbits.  While this seems like a promising outcome, we emphasize that both inwards gas-driven migration \citep{Canup2002, Madeira2021} and long-term outwards tidally-driven migration  \citep{1979Natur.279..767Y,Tittemore1990, Showman1997, Lari2020} have potentially played a role in repositioning the satellites both during and post-formation. Other regions of the CPD parameter space are equally capable of producing solids with \mbox{$0.4 < f_{\rm ice} < 0.55$}, but vary in the position of their snowline. In any case, we believe that it is more meaningful to determine whether a particular CPD can form enough ice on the relevant viscous timescale, rather than at precisely which radii a particular abundance of ice can be found. 

To produce solids with a minimum \mbox{$f_{\rm ice}$ $\sim 0.5$} the global dust-to-gas ratio of the CPD must be $\leq\,10^{-3}$ and does not need to be $< 10^{-3.7}$.  We suggest that this does not represent a minimum $d/g$ limit as imperfect accretion \citep{dwyer2013}, (hydrodynamic) proto-atmospheric escape \citep{Kuramoto1994,Bierson2020}, impact-vaporization \citep{Nimmo2012}, or tidal heating \citep{Canup2009} may all have played a role in reducing the volatile mass of the satellites either during or post-accretion. A minimum $d/g$ is instead implied by the minimum time required to accrete the solid total mass of the Galilean satellites ($\sim 10^{-7}$\,M$_{\odot}$) into the CPD.  We consider the lifetime of the gaseous circumstellar disk ($\sim 10$ Myr) to be the maximum time over which the CPD can continue to accrete gas.  Assuming Jupiter's runaway accretion ended $<3.94$\,Myr after CAI formation \citep{Weiss2021} and that moon formation only began at this stage, this leaves $\sim$6\,Myr to accrete the refractory material for the moons.  We emphasize that this limit is very approximate, as it is possible that circumstellar gas disk lifetimes regularly exceed 10\,Myr \citep{Pfalzner2014}. Conversely an even shorter lifetime ($<4.89$\,Myr after CAI formation) has been proposed for the solar nebula on the basis of paleomagnetic measurements \citep{Weiss2021}.  

The midplane dust-to-gas ratio and thus the $f_{\rm ice}$ in the CPD will differ from what has been derived from our models if the grain size distribution is significantly altered in some way.  The circumstellar disk gap edge may filter out larger grains from the accretion flow \citep{2006MNRAS.373.1619R,Zhu2012}. Grains larger than 100 \textmu m, which settle efficiently, are those primarily responsible for enhancement of the dust mass at the CPD midplane.  Massive planets may however vertically stir circumstellar disk midplane dust outside the gap \citep{Binkert2021}.  \citet{Szulagyi2021} found that significant vertical stirring of large grains occurs at the gap outer wall in the presence of planets with mass above that of Saturn, resulting in a substantial delivery of mm-sized grains to the CPD. The high-mass tail of the dust distribution could alternatively be depleted by the more rapid inwards drift of these grains in the CPD.  We have shown that grains larger than $\sim1$mm no longer advect with the gas in the CPD.  The steady-state dust grain size distribution will thus likely be truncated. In the absence of these grains the limits we have derived on the CPD mass are revised upwards by a factor $\sim\,5$.

\subsection{Does grain drift erase the radial distribution of ices?} \label{sec:discussion_drift}

We have tested only the "gas-starved" disk paradigm in which a CPD must over time accrete the solids to form large moons. The sequential formation, episodic growth, and potential loss of migrating moons is a characteristic of this theory \citep{Canup2002}. In such a dynamical environment the relevancy of the instantaneous radial distribution of icy grains remains to be established. The simplest way in which the chemical properties of dust in the CPD could be imprinted on the final satellite system would be through in-situ formation: the satellites accrete the bulk of their material at fixed radial positions in the CPD \citep{Ronnet2017}. This might occur if the innermost proto-moon were prevented from migrating into Jupiter by the presence of a gas-free magnetically-truncated inner cavity  \citep{Takata1996,Batygin2018,Shibaike2019}.  Additional proto-satellites could then pile-up in a resonant chain and be stabilized against further migration by the proto-moon anchored at the disk inner edge \citep{2009ApJ...699..824O, 2010ApJ...714.1052S,2017MNRAS.470.1750I,Madeira2021}.  Drifting grains would still be free to cross the orbit of proto-moons as accretion efficiency remains low when the proto-moons are only a fraction of their final mass \citep{Shibaike2019}.  In this paradigm proto-moons at relatively fixed positions, continually accreting grains that drift into their feeding-zone \citep{Ronnet2020}.  We find that the ice fraction of small \mbox{(< 1\,cm)} drifting grains in the inner disk will almost always reflect ambient conditions (Sect.  \ref{sec:grain_drift}) independently of whether the gas in the CPD flows radially inwards or outwards (the fate of trapped grains is discussed in Appendix \ref{appendix:trapped-grains}).  If the proto-moons (resonantly anchored at fixed radii within the CPD)  accrete the majority of their total mass from these drifting grains, their bulk ice fraction would reflect the temperature gradient of the CPD.  

\section{Conclusions} \label{sec:conclusions}

Circumplanetary disks represent a unique chemical environment characterized by high-densities and a relatively short timescale on which gas and dust are viscously or aerodynamically lost.  We aimed to explore the process of ice formation in this environment from sharply contrasting initial chemical conditions, knowing that solids with ice/rock $\sim 1$ must be able to form within the viscous timescale of a Jovian CPD. We tested the paradigm in which solids are delivered directly from the circumstellar disk in the form of small grains \mbox{(< 1 cm)} to a "gas-starved", relatively low-mass CPD.  We highlight our key conclusions as follows:

\vspace{4ex}

\textbf{If infalling material is chemically reset:}

\begin{enumerate}

\item High densities in the CPD facilitate three-body ``collider" reactions that lead to rapid water ice formation.  Roughly half of the water ice is formed within a single year by the hydrogenation of OH. 

\item  Solids with the ice fraction of Ganymede or Callisto are produced within $t_{\rm visc}$ for \mbox{$\alpha \approx 10^{-3}-10^{-4}$} if the midplane is depleted in dust by a factor 20-50 relative to the canonical \mbox{$d/g=10^{-2}$}. 

\end{enumerate}

\textbf{If chemical inheritance occurs:} 

\begin{enumerate}

\item  Ices near the planet efficiently sublimate and establish a snowline at a similar location to that of the reset case within $t_{\rm visc}$. Additional ice formation is minimal. 

\end{enumerate}

\textbf{In either case:} 

\begin{enumerate}

\item Icy circumstellar dust grains preserve the majority of their volatile content during gap-crossing if accreted onto the CPD within 100 yr unless the stellar luminosity is \mbox{$>10$ L$_{\odot}$}.  
\item The compositional imprint of the CPD temperature profile is not erased by radial dust drift for grains of size $a < 1$ cm.

\item Only the ``high-mass" CPDs \mbox{($M_{\rm CPD}$ = 10$^{-4}$ M$_{\rm J}$)} are sensitive to the initial chemical conditions:  water ice formation in the inner disk is less efficient if a chemical reset occurs as oxygen tends to remain locked in the gas-phase CO.

\end{enumerate}

In our solar system icy moons are common.  No matter whether or not ices sublimate upon incorporation into the CPD, we have demonstrated that ices can be efficiently re-deposited onto dust grains and enable the general ubiquity of icy moons.

\begin{acknowledgements}
The research of N.O. and I.K. is supported by grants from the Netherlands Organization for Scientific Research (NWO, grant number 614.001.552) and the Netherlands Research School for Astronomy (NOVA). "This research has made use of NASA's Astrophysics Data System Bibliographic Services.  This research has made extensive use of Numpy \citep{numpy},  Matplotlib \citep{matplotlib}, scipy \citep{scipy}, and prodimopy \url{https://gitlab.astro.rug.nl/prodimo/prodimopy}. The authors would like to thank the anonymous referee for comments that contributed to the accuracy, clarity, and focus of this work.
\end{acknowledgements}


\bibliographystyle{aa} % style aa.bst
\bibliography{refs.bib} % your references Yourfile.bib


%
%-------------------------------------------------------------
%          For the appendices, table longer than a single page
%-------------------------------------------------------------

% Table will be print automatically at the end of the document, 
% after the whole appendices

\begin{appendix} %First appendix

\section{The RT diffusion solver} \label{appendix:rtdiffusionsolver}

\def\kRoss{\kappa_{\rm R}}
\def\kPl{\kappa_{\rm P}}
\def\div{{\rm div}}
\def\grad{{\bf grad}}
\def\half{\hspace*{0.2pt}\sfrac{1\hspace*{-0.4pt}}{\hspace*{0.2pt}2}}
\def\ver{{\rm ver}}
\def\hor{{\rm hor}}

{\sc ProDiMo} solves the radiative transfer equation \citep[see Eq.\,(12)
  in][] {Woitke2009a} together with the condition of radiative energy
conservation, which in general can be written as
\begin{equation}
  \div \vec{F} ~=~ \Gamma \ ,
  \label{eq:divF}
\end{equation}
where $\vec{F}=\int \vec{F}_\nu\,d\nu\rm\;[erg/cm^2/s]$ is the
bolometric radiation flux vector and $\Gamma\rm\,[erg/cm^3/s]$ is the
non-radiative heating rate per volume.  In the viscous case, we use
$\Gamma\!=\!\Gamma_{\rm vis}$, see Eq.\,(\ref{eq:visc_heating}) with stellar mass $M_\star$ and
stellar radius $R_\star$ instead of $M_{\rm p}$ and $R_{\rm p}$ for circumstellar
discs.  The additional non-radiative heating leads to a surplus
emission of photon energy according to
\begin{equation}
  4\pi \int\kappa_\nu^{\rm abs}\,\Big(B_\nu(T)-J_\nu\Big)\,d\nu ~=~ \Gamma \ ,
  \label{eq:radEq}
\end{equation}
where $\kappa_\nu^{\rm abs}$ is the dust absorption coefficient at
frequency $\nu$, $B_\nu(T)$ the Planck function, and $J_\nu$ the
mean intensity. For passive discs, we have $\Gamma\!=\!0$, in which case
Eq.\,(\ref{eq:radEq}) simplifies to the ordinary condition of
radiative equilibrium.

The numerical solution method for the radiative transfer (RT) in {\sc
  ProDiMo} involves iterations where formal solutions with isotropic
scattering are computed along multiple rays to cover the full $4\pi$ solid
angle as seen from every point in the disc, see Sect. ~4 in
\citet{Woitke2009}. A formal solution results in new $J_\nu(r,z)$
which are used to update the dust temperatures $T_{\rm dust}(r,z)$ and
source functions.  The convergence of this $\Lambda$-iteration is
accelerated by the Ng-method \citep[see][]{Auer1984}. However, despite this
acceleration, the convergence is still slow in the midplane, which
is a serious problem for all radiative transfer codes for discs,
including the Monte-Carlo codes, see \citet{Pinte2009}.

Here we describe a method how to avoid this problem.
In the diffusion approximation, the bolometric radiation flux 
\begin{equation}
  \vec{F}(r,z) = -\frac{4\,\pi}{3\kRoss(r,z)}\,\grad\,J(r,z)
\end{equation}
is given by the gradient of the bolometric mean intensity $J(r,z)$.
The Rosseland-mean and Planck-mean opacities are defined as
\begin{align}
  \frac{1}{\kRoss(r,z)} =&~
    \int \frac{1}{\kappa^{\rm ext}_\nu(r,z)} \frac{dB_\nu(T)}{dT}\,d\nu \;\;\Big/\;
    \int \frac{dB_\nu(T)}{dT}\,d\nu \\
  \kPl(r,z) =&~
    \int \kappa^{\rm abs}_\nu(r,z) B_\nu(T)\,d\nu \;\;\Big/\;
    \int B_\nu(T)\,d\nu \ ,
\end{align}
where $\kappa^{\rm ext}_\nu(r,z)$ is the extinction coefficient and
$T=T_{\rm dust}(r,z)$ the dust temperature at position $(r,z)$ in the disk.

\begin{figure}[!t]
  \centering
  \includegraphics[width=90mm,trim=90 45 50 100,clip]{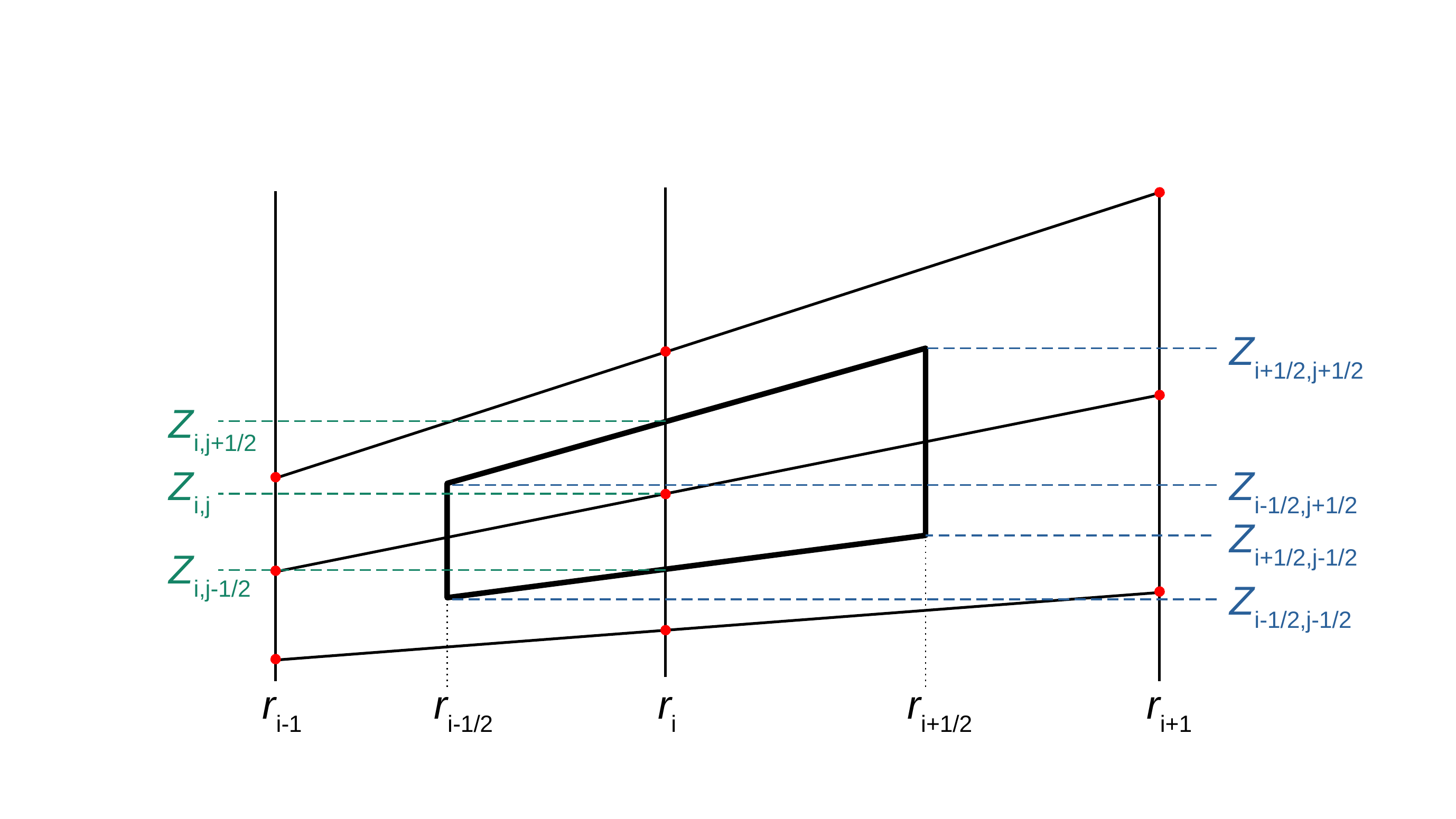}
  \caption{Volume and areas for the spatial cell around point
    $(i,j)$. The mean intensities $J$ and Rosseland opacities $\kRoss$
    are available on the grid points $(i,j)$ marked with red dots.}
  \label{fig:grid}
\end{figure}

At the beginning of a new RT iteration, the Rosseland and Planck
opacities are calculated based on the frequency-dependent disk opacity
structure and the current $T_{\rm dust}(r,z)$. Next, we compute radial
and vertical Rosseland optical depths $\tau_{\rm Ross}=\int
\kRoss\,ds$ from every point. When the radial inward, radial outward
and vertical upward Rosseland optical depths from that point are all
larger than a threshold value (we use a value of 10 here), the point
is flagged as being optically thick, and added to the subset of optically
thick points
\begin{equation}
  {\cal M} = \{\,(i,j)\;|\;(r_i,z_{i,j})\rm\ is\ optically\ thick\} \ .
\end{equation}
where $i$ and $j$ are the 2D-indices of a grid point at radius $r_i$
and height $z_{i,j}$. The following method only updates the mean
intensities $J_{i,j}$ and dust temperatures $T_{i,j}$ on the optically
thick grid points $(i,j)\in\!{\cal M}$, whereas all
other points are regarded as fixed boundary conditions for this
problem. To pick up the bolometric mean intensities on the boundary
points, we integrate Eq.\,(\ref{eq:radEq}) assuming that $J_\nu$ is
close to a Planckian, hence
\begin{equation}
  J(r,z) = B(T)-\frac{\Gamma(r,z)}{4\pi\,\kPl(r,z)} \ .
  \label{eq:JJ}
\end{equation}
where $B(T)\!=\!\sigma\,T_{\rm dust}(r,z)^4/\pi$ is the
frequency-integrated Planck function. Integration of
Eq.\,(\ref{eq:divF}) over the volume associated with grid point
$(i,j)$ as sketched in Fig.~\ref{fig:grid} results in the following
numerical equation
\begin{align}
 & A^\ver_{i-\half,j} D_{i-\half,j} \frac{J_{i,j}-J_{i-1,j}}{\Delta r_{i-\half}}
 ~+~A^\ver_{i+\half,j} D_{i+\half,j} \frac{J_{i,j}-J_{i+1,j}}{\Delta r_{i+\half}}
  \label{eq:balance}\\
 & +~A^\hor_{i} D_{i,j-\half} \frac{J_{i,j}-J_{i,j-1}}{\Delta z_{i,j-\half}}
 ~+~A^\hor_{i} D_{i,j+\half} \frac{J_{i,j}-J_{i,j+1}}{\Delta z_{i,j+\half}}
 ~=~ V_{i,j}\,\Gamma_{i,j} \nonumber \ ,
\end{align}
where we note that the vertical fluxes through the cell boundaries
involve a scalar product with the slanted normal vector of the surface
area, hence $A^\hor_{i}$ is the cell's horizontal area after
projection onto the vertical direction. The following abbreviations
are used for the distances, vertical and horizontal areas, and
the volume.  They are given by the geometry of the {\sc ProDiMo} grid
points, which are aligned on radial rays on which $z/r$ is constant
\begin{align}
  r_{i-\half}  =&~\sqrt{r_i r_{i-1}} \label{eq:first}\\
  r_{i+\half}  =&~\sqrt{r_i r_{i+1}} \\
  \Delta r_{i-\half} =&~r_i-r_{i-1}\\
  \Delta r_{i+\half} =&~r_{i+1}-r_i\\
  z_{i,j-\half} =&~\frac{1}{2}(z_{i,j}+z_{i,j-1}) \\ 
  z_{i,j+\half} =&~\frac{1}{2}(z_{i,j}+z_{i,j+1}) \\ 
  z_{i-\half,j-\half} =&~z_{i,j-\half} \frac{r_{i-\half}}{r_i}\\
  z_{i-\half,j+\half} =&~z_{i,j+\half} \frac{r_{i-\half}}{r_i}\\
  z_{i+\half,j-\half} =&~z_{i,j-\half} \frac{r_{i+\half}}{r_i}\\
  z_{i+\half,j+\half} =&~z_{i,j+\half} \frac{r_{i+\half}}{r_i}\\
  A^\ver_{i-\half,j} =&~2\pi\,r_{i-\half}(z_{i-\half,j+\half}-z_{i-\half,j-\half})\\
  A^\ver_{i+\half,j} =&~2\pi\,r_{i+\half}(z_{i+\half,j+\half}-z_{i+\half,j-\half})\\
  A^\hor_{i} =&~\pi(r_{i+\half}^2-r_{i-\half}^2)\\
  V_{i,j} =&~ A^\hor_{i}(z_{i,j+\half}-z_{i,j-\half})
\end{align}
The radiative diffusion coefficients are defined as
\begin{align}
  D_{i,j} =&~ \frac{4\pi}{3\kRoss(r_i,z_{i,j})}\\
  D_{i-\half,j} =&~ \sqrt{D_{i,j} D_{i-1,j}} \\
  D_{i+\half,j} =&~ \sqrt{D_{i,j} D_{i+1,j}} \\
  D_{i,j-\half} =&~ \sqrt{D_{i,j} D_{i,j-1}} \\
  D_{i,j+\half} =&~ \sqrt{D_{i,j} D_{i,j+1}} \label{eq:last}
\end{align}
Equation~(\ref{eq:balance}) states a system of linear equations for the
unknown bolometric mean intensities $J_{i,j}$ on the optically thick points
$(i,j)\in{\cal M}$ of the form
\begin{equation}
  {\cal A}\cdot \vec{X} = \vec{B}  \ ,
  \label{eq:matrix}
\end{equation}  
where all quantities in Eqs.\,(\ref{eq:first}) to (\ref{eq:last}) are
constants forming the matrix $\cal A$, and the volumes $V_{i,j}$
and heating rates $\Gamma_{i,j}$ are constants forming the rest vector
$\vec B$.  The unknowns $\{J_{i,j}\}$ at the optically thick points
$(i,j)\in{\cal M}$ constitute the solution vector $\vec{X}$.  All
terms in Eq.\,(\ref{eq:balance}) that involve the other $J_{i,j}$
on the adjacent points are also included into $\vec{B}$. The matrix
equation to solve (Eq.\,\ref{eq:matrix}) has a typical dimension
of a few hundred to a few thousand, depending on disk mass, geometry,
and dust parameters.  

This way we can solve the 2D radiative diffusion problem for the
unknown mean intensities in the optically thick region as a linear
boundary value problem in one go, where there is one layer of points
surrounding the optically thick regions which sets the boundary
values. Our method calculates how the disk transports the photon
energy through the optically thick core inside of the boundary layer.
It is applicable to both cases, passive discs without viscous
heating and active discs with $\Gamma\!>\!0$.

Once the $\{J_{i,j}\}$ on $(i,j)\in{\cal M}$ have been determined, we
revert the process described by Eq\,(\ref{eq:JJ}) 
\begin{align}
  B(T) =&~J(r,z) + \frac{\Gamma(r,z)}{4\pi\,\kPl(r,z)}  \\
  T_{\rm dust}(r,z) =&~ \left(\frac{\pi}{\sigma}B(T)\right)^{1/4}
\end{align}
and multiply the frequency-dependent mean intensities $J_\nu(r,z)$, as
they were determined prior to the application of the diffusion solver,
by a constant factor to make Eq.\,(\ref{eq:radEq}) valid again,
thereby keeping the previously calculated frequency distribution of
$J_\nu(r,z)$.

After having modified $T_{\rm dust}(r,z)$ and $J_\nu(r,z)$ this way on
all grid points $(i,j)\in{\cal M}$, the normal RT solution method
resumes, which begins by calculating the source functions on all
grid points and continues by performing a formal solution.

\begin{figure}[!th]
  \includegraphics[width=90mm]{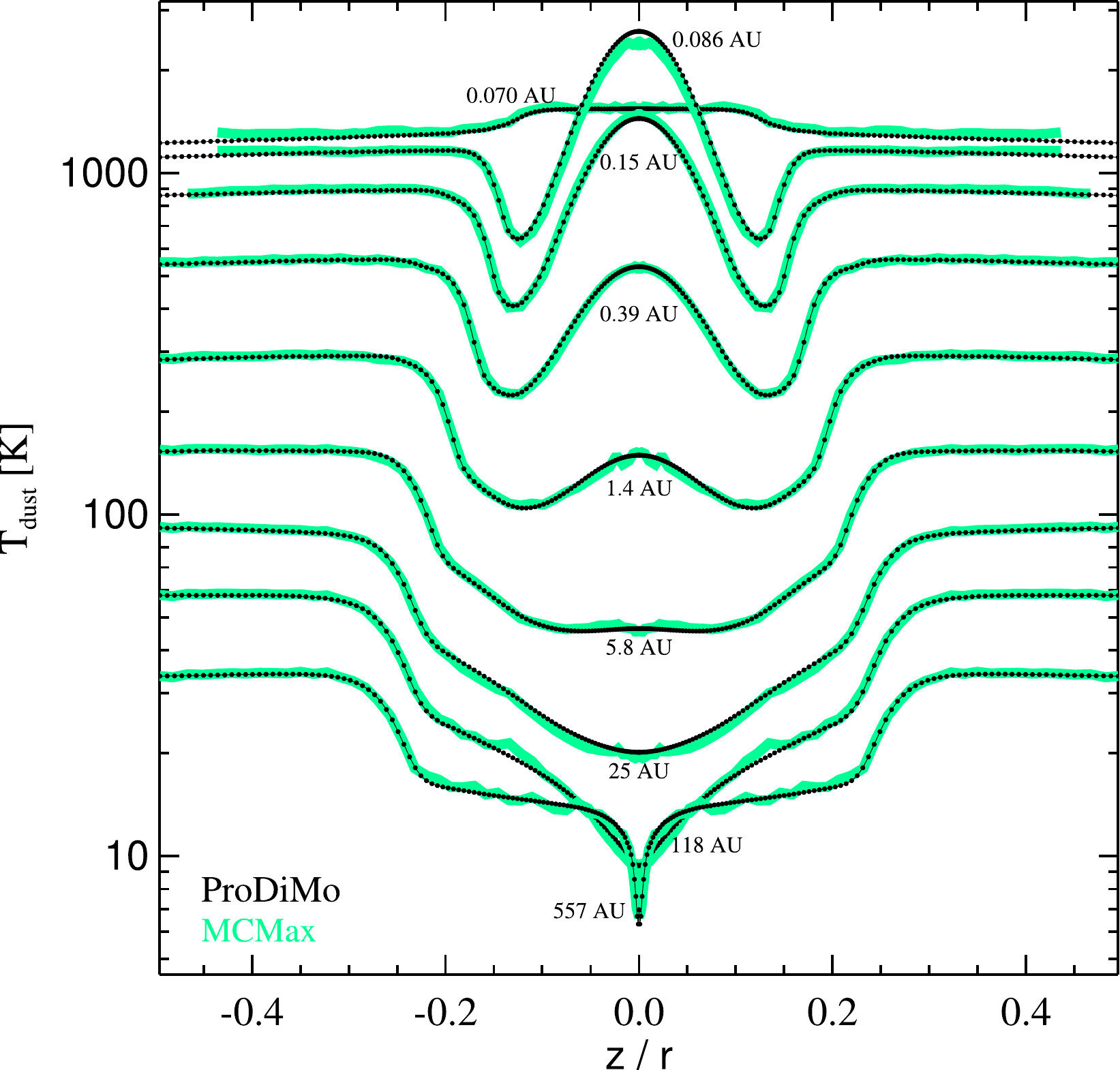}
  \caption{Benchmark test for a viscous circumstellar disk with a disk
    mass of 0.01\,$M_\odot$ and a mass accertion rate of $\dot{M}_{\rm
      acc}\!=\!10^{-8}\rm\,M_\odot/yr$, see text. The green lines are
    temperature cuts at selected radii calculated by MCMax, the small
    wiggles are due to the Monte-Carlo noise. The black
    dots show the temperatures calculated by {\sc ProDiMo}.}
  \label{fig:benchmark1}
\end{figure}

Figure~\ref{fig:benchmark1} shows a benchmark test against the Monte
Carlo radiative transfer program MCMax \citep{Min2011}. We
consider the disk model that is described in detail by
\citet{Woitke2016}, see their table~3. The central star is a 2\,Myr
old T\,Tauri star with a mass of 0.7\,$M_\odot$ and a luminosity of
1\,$L_\odot$, the disk has a mass of 0.01\,$M_\odot$, with a gas/dust
ratio of 100. The dust is composed of 60\% silicate, 15\% amorphous
carbon and 25\% porosity. The dust grains have sizes between
0.05\,$\mu$m and 3\,mm, with an unsettled powerlaw size-distribution
of index -3.5. The dust is settled according to the prescription of
\citet{Dubrulle1995} with $\alpha_{\rm settle}\!=\!0.01$. In contrast
to this standard passive T\,Tauri model, we use here a mass accretion
rate of $\dot{M}_{\rm acc}\!=\!10^{-8}\rm\,M_\odot/yr$ to set the
viscous heating of the disk according to Eq.\,(\ref{eq:visc_heating}). We use $140$
radial $\times100$ vertical grid points, and 40 wavelength bins.
Figure~\ref{fig:benchmark1} shows good agreement.

\begin{figure}[!b]
  \centering
  \includegraphics[width=70mm,trim=0 0 0 20,clip]{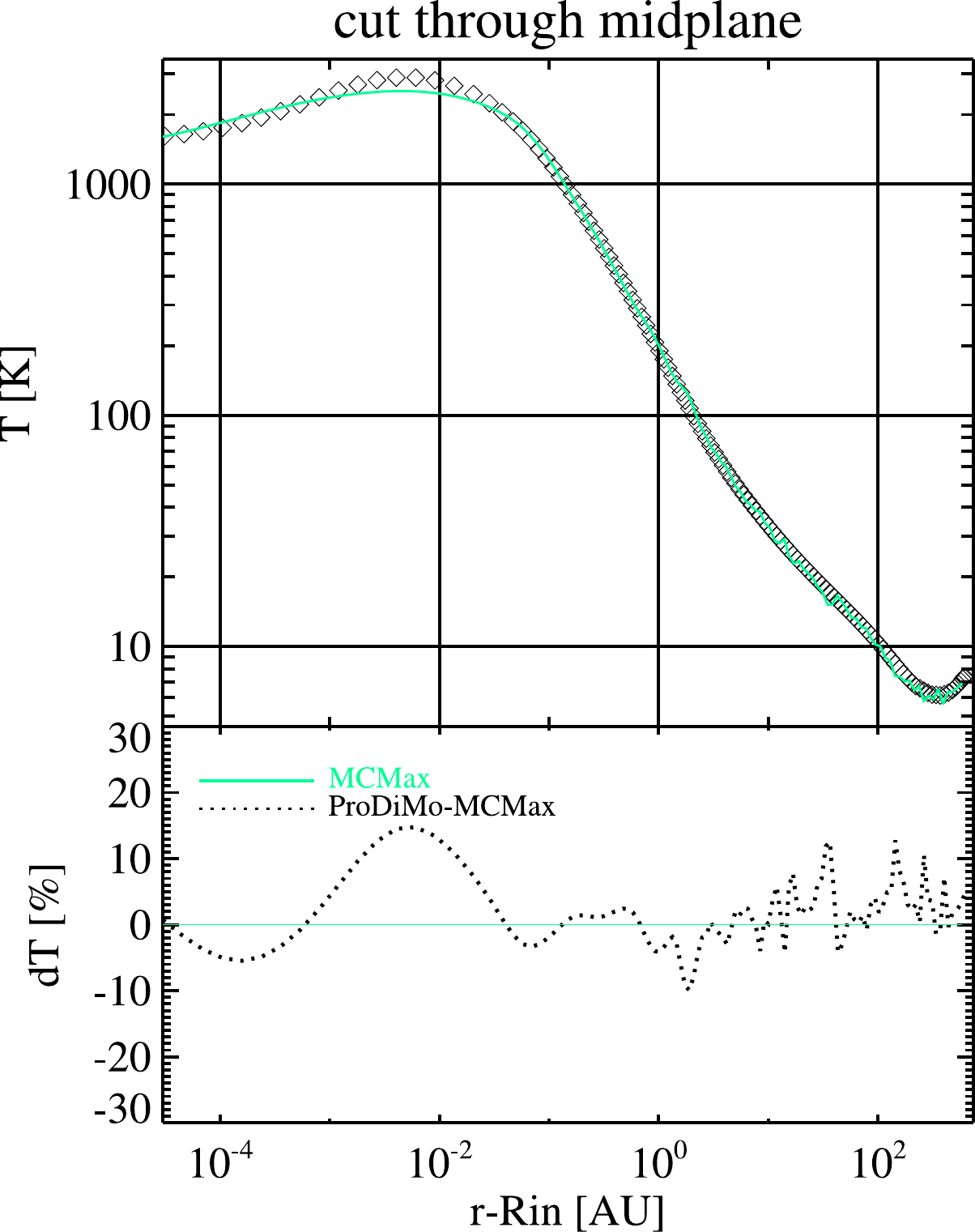}
  \caption{Midplane temperature $T_{\rm dust}(r,0)$ as function of the
    log-distance from the inner rim. Again, the green line is 
    the temperature profile calculated by MCMax, and the small
    black diamonds show the temperature values calculated
    by {\sc ProDiMo}.}
  \label{fig:benchmark2}
\end{figure}

\noindent Figure~\ref{fig:benchmark1} reveals a number of interesting
features in the disk temperature structure:
\begin{itemize}
  \item The dust temperature
    at the inner rim is not much affected by viscous heating.
  \item From top to midplane, the temperature first decreases in the
    disk shadow, but then the trend is reversed and the temperature
    re-increases towards the midplane as the viscous heat pumped into the
    disk needs to flow outward, that is mostly upward, which according
    to the diffusion approximation requires a negative temperature
    gradient.
  \item There is little effect of viscous heating on $T_{\rm dust}$
    outside of the optically thick region which extends outward to
    about 10\,au and upward to about $z/r \approx 0.1-0.15$ in this
    model.
  \item The temperature profile across the midplane beyond the
    tapering radius \citep[$R_{\rm tap}\!=\!100\,$au, see][]{Woitke2016} shows a deep minimum around the midplane
    $z\!=\!0$.  This is because of the extreme dust settling that
    occurs in these diluted outer disk regions, creating more optical
    thickness along the midplane, which brings down the
    midplane temperature to only 6\,K in this model.
\end{itemize}

\noindent Figure~\ref{fig:benchmark2} compares the calculated midplane
temperatures between {\sc ProDiMo} and MCMax, which reveals dust temperatures as high as 2800\,K,
which is of course questionable because at such temperatures, the dust
grains are expected to sublimate.   

\begin{figure}[H]
  \centering
  \includegraphics[width=70mm,trim=30 15 20 20,clip]{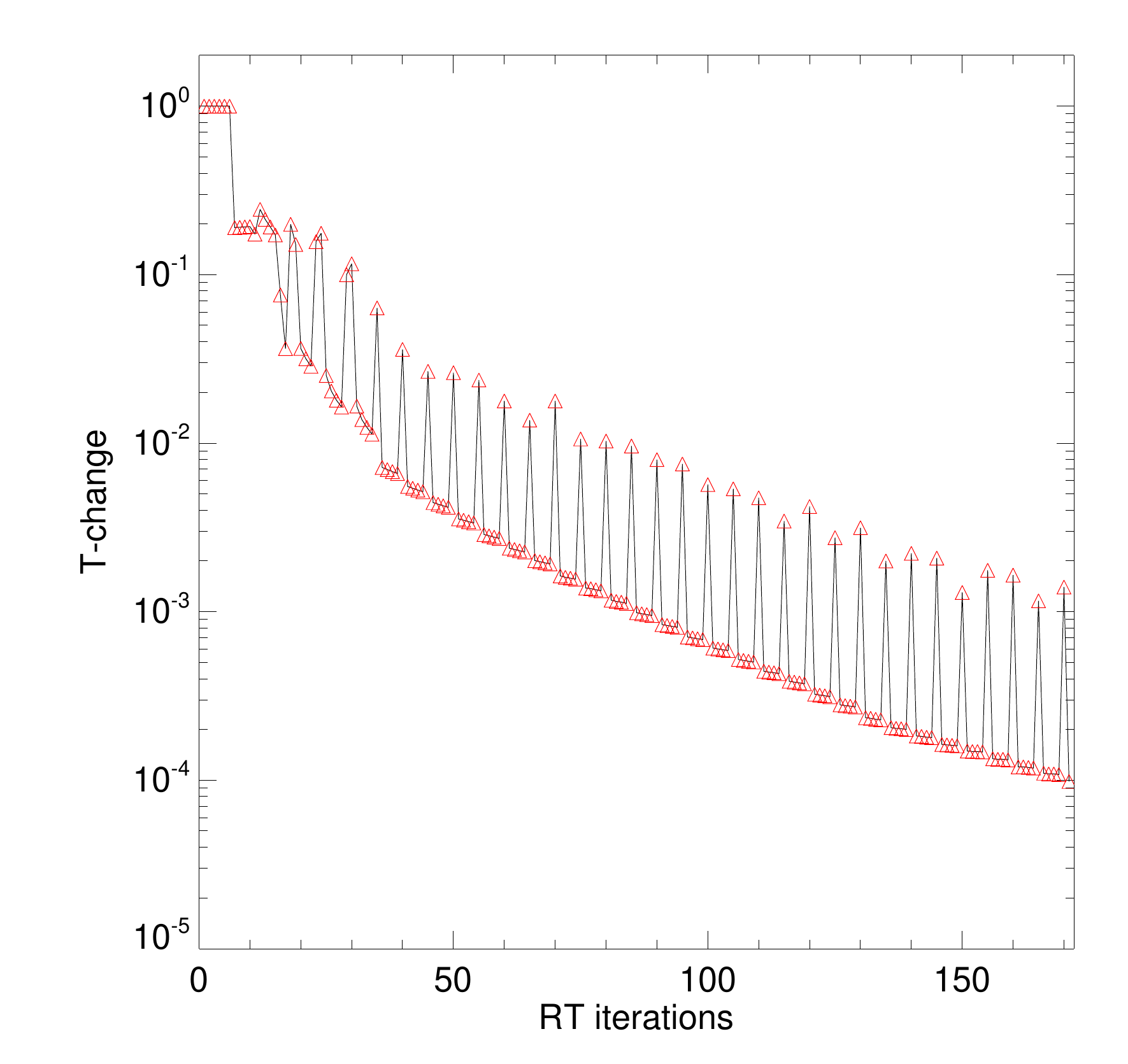}
  \caption{Maximum relative temperature change between iterations as
    function of RT iteration number.}
  \label{fig:benchmark3}
\end{figure}

\noindent Figure~\ref{fig:benchmark3} shows the convergence of the RT
method, achieving residual relative temperature changes smaller than
$10^{-4}$ after about 150 RT iterations.  The maximums occuring each
5$^{\rm th}$ iteration are due to the Ng-acceleration algorithm.

\vfill

\section{Adsorption energies} \label{appendix:eads}

%\begin{enumerate}
%    \item H2O   : 5700 K
%    \item NH3   : 3800 K 
%    \item CO2   : 2580 K 
%    \item CO    : 899  K
%    \item CH4   : 1300 K
%    \item NH2   : 3200 K
%    \item CH3OH : 5530 K
%\end{enumerate}

\begin{table}[H]
\caption{Adsorption energies of the most prevalent molecular ices found in our model CPDs.}

\centering
\renewcommand{\arraystretch}{1.1}%
\begin{tabular}{lll}

\hline \hline 
Ice & E$_{\rm ads}$ [K]  & ref. \\ \hline
H$_2$O     & 4800       &  a        \\
NH$_3$     & 5534       &  b     \\
NH$_2$     & 3956       &  b     \\
C$_2$H$_6$ & 2300       &  c     \\  
C$_2$H$_4$ & 3487       &  b     \\
C$_2$H$_2$ & 2587       &  d     \\
CO$_2$     & 2990       &  e     \\
CH$_3$OH   & 5534       &  a     \\
OH         & 2850       &  b     \\   \hline

\label{tab:Eads}
\end{tabular}
\\
\raggedright

\footnotesize{(a) \citep{Brown2006}} \\
\footnotesize{(b) \citep{Garrod2006}} \\
\footnotesize{(c) \citep{Oberg2009}} \\
\footnotesize{(d) \citep{Collings2004}} \\
\footnotesize{(e) \citep{McElroy2013}} \\

%\footnotesize{(b) \citep{Aikawa1996a}} \\
%\footnotesize{(c) \citep{Girardet2001}} \\
%\footnotesize{(d) \citep{Wakelam2017}} \\

% OH         & 2850       &  b      \\
% HNO        & 2050       &  a        \\

\end{table}

The adsorption energies of our most common ices and their respective references are listed in Table \ref{tab:Eads}.

\vfill
\section{Surface density slope} \label{appendix:surfacedensityslope}

\begin{figure}[H]
  \includegraphics[width=\textwidth/2]{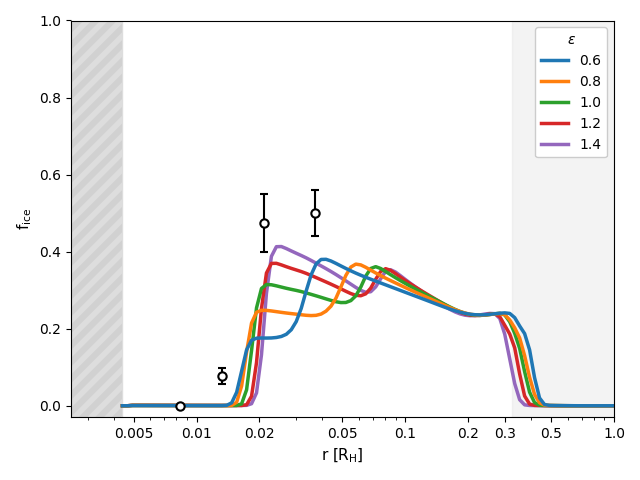}
      \caption{Midplane ice mass fraction for a variety of surface density powerlaw exponents $\epsilon$ for the (7-11) reset CPD with global $d/g = 10^{-3.3}$.}
      \label{fig:appendix-epsilon}
\end{figure}

Our reference CPDs have a surface density powerlaw exponent $\epsilon = 1$.  The steady-state solution for a constant-$\dot M$ decretion $\alpha$-disk is $\epsilon \approx 1.25$ \citep{Batygin2020}.  The midplane ice mass fraction for a variety of possible values of $\epsilon$ is shown in Fig. \ref{fig:appendix-epsilon}.  For a higher $\epsilon$  the NH$_3$ iceline responsible for the ``bump" in the $f_{\rm ice}$ profile at 0.07 R$_{\rm H}$ moves outwards only negligibly.  In the inner disk however the ice mass fraction increases due to a combination of the lower midplane dust-to-gas and more efficient H$_2$O formation.

\vfill
\section{Background temperature}

Throughout this work we have assumed that the CPD is embedded in a radiation field in which the equilibrium dust temperature is 50 K. The midplane dust temperature at the gap center within the circumstellar disk is 50$\pm2$ K, for  a solar luminosity 0.83 L$_{\odot}$, gap A$_{\rm V}$ = 0.008, and heliocentric distance 5.2 au.  For earlier formation times with correspondingly higher solar luminosities (2.34-13.6 L$_{\odot}$), we find gap midplane dust temperatures ranging from 70-120 K at 5.2 au.  

\begin{figure}[H]
  \includegraphics[width=\textwidth/2]{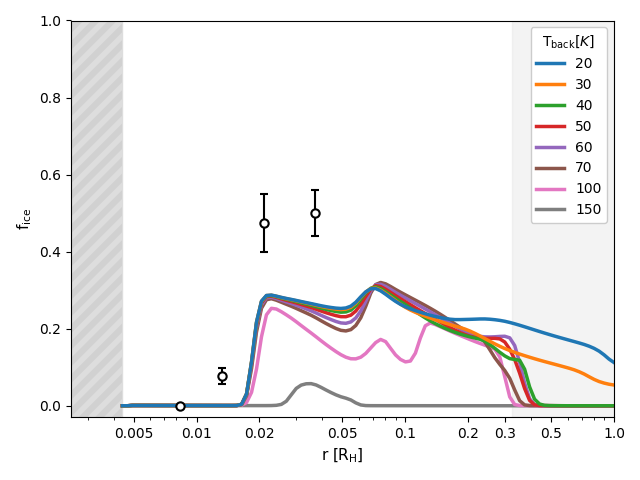}
      \caption{Midplane ice mass fraction for a variety of background temperatures $T_{\rm back}$  for the (7-11) reset CPD and global $d/g = 10^{-3.3}$. The four empty circles with error bars represent the Galilean satellites at their present day locations and composition. The gray striped region on the left represents the inner cavity and the light gray shaded region on the right represents the gravitationally unstable zone.}
      \label{fig:appendix-tback}
\end{figure}

We have assumed that the final stage of Jupiter's accretion and moon formation occured at a radial distance from the sun of 5.2 au.  Volatile enrichment in Jupiter's atmosphere indicates it may have formed further out at circumstellar disk temperatures < 25 K or at radii > 30 au \citep{Oberg2019}.  The Nitrogen abundance of Jupiter, approximately 4$\times$ solar, may suggest additional N$_2$ was accreted from the solar nebula near the N$_2$ iceline \citep{Bosman2019}. In light of this possibility we consider also lower background temperatures down to 20 K.    The midplane $f_{\rm ice}$ for the reference (7-11) CPD can be seen in Fig.\ref{fig:appendix-tback}. Inside the optically thick region of the CPD the influence of the background temperature $T_{\rm back}$ is marginal for temperatures \mbox{$\leq$ 70 K}.  Above 70 K the more volatile NH$_3$ and CO$_2$ are unstable as ices and only water ice remains.  Below 40 K the outer disk is able to retain ices at radii where A$_{\rm V} < 1$ as the photodesorption timescale is in excess of the viscous timescale.

%%%%%%%%%%%%%%%%%%%%%%%%%%%%%%%%%%%%%%%%%%%%%%%%%%%%%%%%%%%%%%%%
\vfill
\section{Vertical mixing} \label{appendix:vertical-mixing}

\begin{figure}[H]
  \includegraphics[width=\textwidth/2]{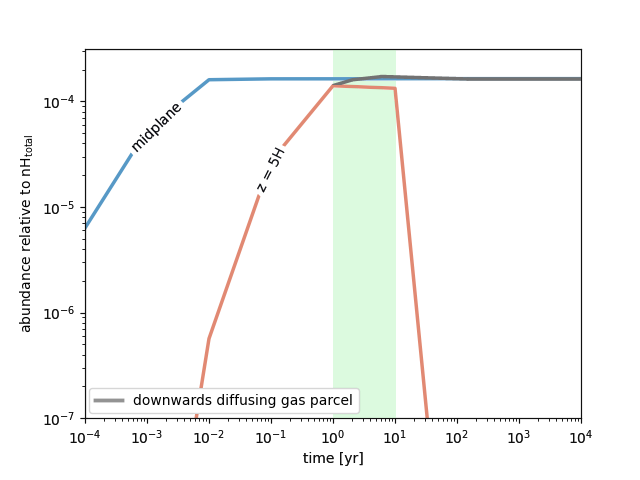}
          \caption{Evolution of the water ice abundance relative to the total number of hydrogen nuclei at the midplane (blue) and at 5 scale heights above the CPD midplane (red), and in the case of the gas parcel which drifts from $z=5H$ downwards to the midplane over a period of 100 yr (grey).  The period in which the altitude of the gas parcel was iteratively lowered towards the midplane is highlighted in light green.}
      \label{fig:downwards-diffuse}
\end{figure}

We have made the simplifying assumption that material which accretes onto the CPD is instantaneously distributed vertically throughout the disk. The shock front may be found at a few ($\sim 5$) scale heights above the CPD midplane \citep{Tanigawa2012}. At 5 scale heights above the centrifugal radius the dust temperature $T_{\rm dust}$ = 123 K (relative to 89.5 K at the midplane), and optical extinction A$_{\rm V}$=0.004 (16.4 at the midplane).  The velocity of vertical mixing by turbulent diffusion can be estimated as $v_{\rm z} = \alpha c_{\rm s}$ where $c_{\rm s}$ is the local speed of sound \citep{Heinzeller2011}.  We find $v_{\rm z} \sim 0.5-1$ m s$^{-1}$ in this region for the high-viscosity CPDs, assuming that the magnitude of the turbulence is constant from the midplane up to $z = 5H$.  The resulting vertical diffusion mixing timescale is $0.01$ $t_{\rm visc}$ (10-100 yr).  We perform a test in the (7-11) CPD in which a parcel of gas is accreted at $z = 5H$ and iteratively evolve its chemistry in steps as it diffuses towards the midplane over 10 yr to understand the impact of more tenuous and high temperature conditions in the initial stages of ice formation (see Fig. \ref{fig:downwards-diffuse}).  By the end of the stage of rapid formation of water (1-2 yr), the ice abundance has equalized to the fiducial case at the midplane.

\vfill
\section{Continued ice deposition on trapped grains} \label{appendix:trapped-grains}

In each of the reference CPDs grains of a certain size range remain trapped within the CPD if we assume that gas actively decretes via the outer edge of the disk.  A trapped grain will increase physically in size as ice adsorbs onto its surface and thus alter its aerodynamic properties.  \citet{Batygin2020} proposes that grain growth and sublimation could play a role in trapping and radial cycling of grains with size 0.1-10 mm. The size range of trapped grains is $\sim0.01-1$ mm in our high-mass CPDs and $\sim10^{-3}-0.01$ mm in the low-mass CPDs, representing 2.5$\%$ and 0.1$\%$ of the infalling dust mass that reaches the midplane, respectively.   A modal icy grain is typically coated in no more than 4000 monolayers of water ice.  Assuming a monolayer thickness $\sim0.5$ nm  \citep{Zangi2003} and compact morphology, an icy mantle no more than 1 \textmu m thick will form.  A grain of size 0.05 \textmu m can thus increase in size by a factor 20 and have a density corresponding more closely to water ice rather than silicate.  For a 1 \textmu m thick mantle the aerodynamic effect of increased cross-section is negated by the corresponding reduction in grain density.  If the trapped grain icy mantles continue to grow mantles beyond several 1000 monolayers, the new equilibrium trap radius drifts inwards.  Realistically only a fraction (0.01-0.1$\%$) of the total CPD gas mass is accreted per year.  Mantle growth for a trapped grain will thus not exceed $\sim2$ nm yr$^{-1}$ on average. The time for a grain to grow an icy mantle that allows it to drift to the trap inner edge is then $\sim10^6$ yr (high-mass CPD) or $\sim10^5$ yr (low-mass CPD) assuming a compact grain structure. Ice deposition is thus unlikely to allow grains to escape traps. This estimate does not take into consideration  grain growth by coagulation or  fragmentation by collision.

\section{Chemical abundances of the 0D "molecular cloud" model} \label{sec:appendix:mc}

\begin{figure}[H]
  \includegraphics[width=\textwidth/2]{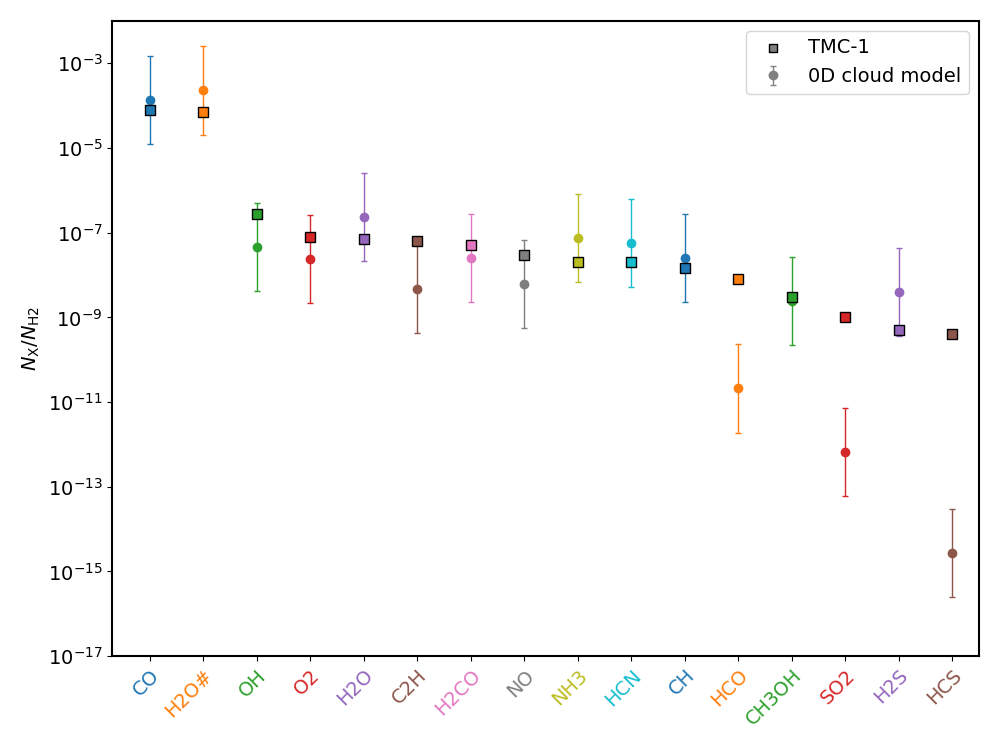}
      \caption{Ratio of the column density of several common species  relative to H$_2$ in the 0D molecular cloud model (circles with error bars)  to observational values for TMC-1 (squares with black border).  Error bars on model values represent a factor 10x uncertainty for illustrative purposes.  TMC-1 abundances are taken from \citep{Suutarinen2011} (OH,CH), \citep{Sakai2010} (C$_2$H), \citep{Smith2004, Walsh2009} otherwise.}
      \label{fig:appendix-mc}
\end{figure}

The input parameters of the 0D molecular cloud model can be found in Table \ref{tab:mol_cloud}.  A comparison between the model column densities of several common species with observations of TMC-1 can be found in Fig. \ref{fig:appendix-mc}.  While most of the common species column density agrees relatively well with observations, the abundance of S-bearing species hinge on the  uncertainties regarding the S elemental abundance \citep{Ruffle1999}.

\begin{table}[H]
\caption{Molecular cloud parameters}
\centering
\setlength{\tabcolsep}{1pt}
\begin{tabular}{llll}
 \hline  \hline  
Parameter          & Value               & Unit          \\   \hline
Hydrogen density   &    $10^4$           & cm$^{-3}$     \\  
Gas temperature    &    10.0             & K             \\
Dust temperature   &    10.0             & K             \\ 
Optical extinction &    10.0             &  -            \\ 
Mean grain radius  &    0.1              & \textmu m     \\
Cloud Lifetime     &    $1.7\times 10^5$ & yr \\

\end{tabular}

\caption*{\textbf{Note}: Parameter values of the molecular cloud model and integration time are chosen according to the method of \citet{Helling2014} recommended for TMC-1 (Taurus Molecular Cloud) by \citet{McElroy2013}. Initial atomic abundances intended to represent typical diffuse interstellar medium conditions are also adopted from \citep{McElroy2013}.}

\label{tab:mol_cloud}
\end{table}

\section{CPD dust-to-gas ratio} \label{appendix:dusttogas}

\begin{figure*}

\centering
\begin{subfigure}{\textwidth/2}
    \centering
  \includegraphics[width=\textwidth]{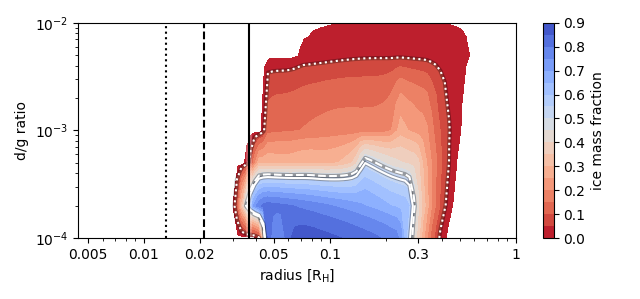} %DG-609
     \caption{High-Mass High-Viscosity (7-10) reset }
\end{subfigure}%
\begin{subfigure}{\textwidth/2}
    \centering
  \includegraphics[width=\textwidth]{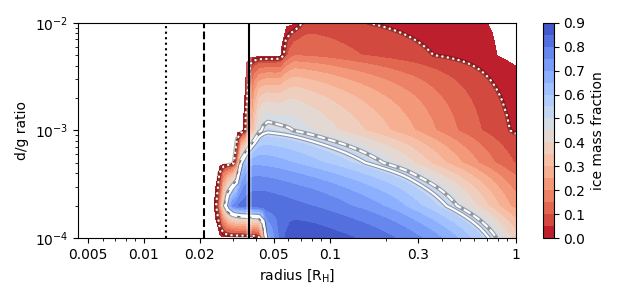}
     \caption{High-Mass High-Viscosity (7-10) inherit }
\end{subfigure}
\begin{subfigure}{\textwidth/2}
    \centering
  \includegraphics[width=\textwidth]{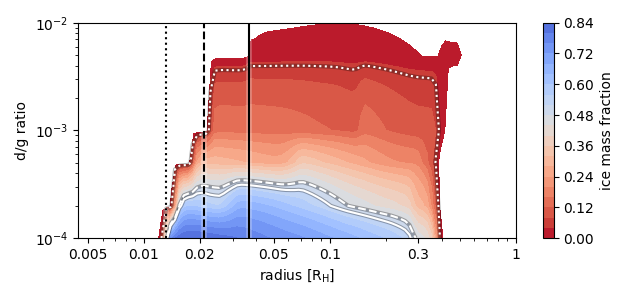} %DG-609
     \caption{High-Mass Low-Viscosity (7-11) reset }
\end{subfigure}%
\begin{subfigure}{\textwidth/2}
    \centering
  \includegraphics[width=\textwidth]{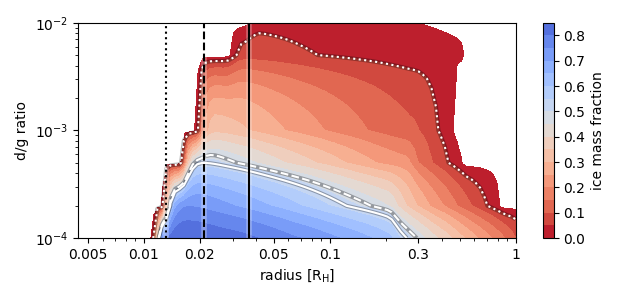}
     \caption{High-Mass Low-Viscosity (7-11) inherit }
\end{subfigure}
\begin{subfigure}{\textwidth/2}
    \centering
  \includegraphics[width=\textwidth]{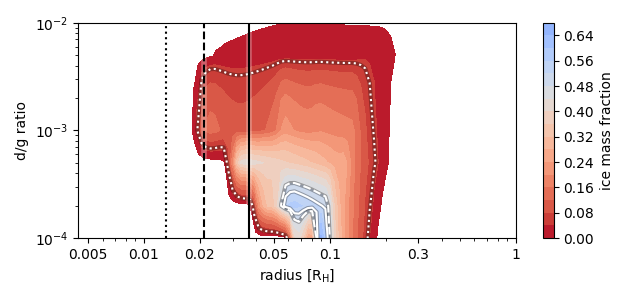} %DG-609
     \caption{Low-Mass High-Viscosity (8-11) reset}
\end{subfigure}%
\begin{subfigure}{\textwidth/2}
    \centering
  \includegraphics[width=\textwidth]{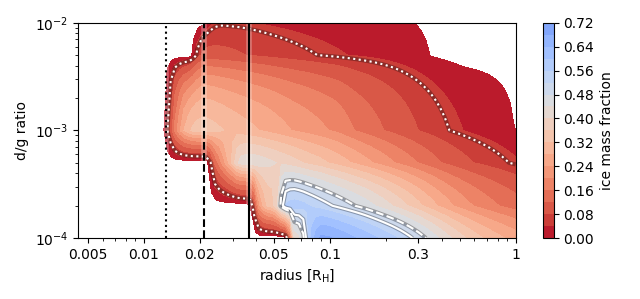}
     \caption{Low-Mass High-Viscosity (8-11) inherit}
\end{subfigure}
\begin{subfigure}{\textwidth/2}
    \centering
  \includegraphics[width=\textwidth]{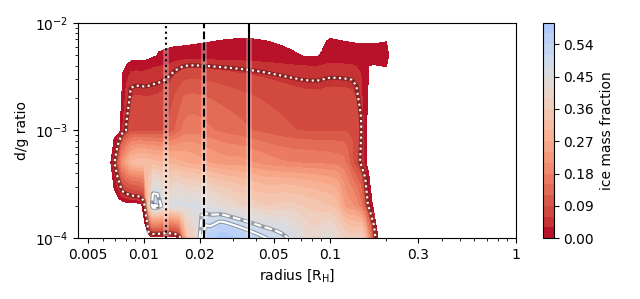} %DG-609
     \caption{Low-Mass Low-Viscosity (8-12) reset}
\end{subfigure}%
\begin{subfigure}{\textwidth/2}
    \centering
  \includegraphics[width=\textwidth]{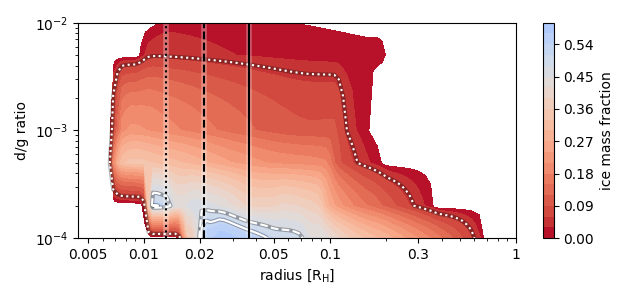}
     \caption{Low-Mass Low-Viscosity (8-12) inherit}
\end{subfigure}

\caption{Midplane radial ice mass fraction $f_{\rm ice}$ of the dust-to-gas ratio parameter exploration for the chemically reset (left column) and chemically inherited (right column) reference CPDs.  The black dotted, dashed, and solid lines indicate the radial position of Europa, Ganymede, and Callisto.  The white dotted, dashed, and solid lines indicate where $f_{\rm ice}$ is equivalent to the estimated ice mass fraction of Europa (0.05), Ganymede (0.48), and Callisto (0.52).}
\label{fig:fice_dustgas_001}
\end{figure*}

In Fig. \ref{fig:fice_dustgas_001} we explore the resulting midplane ice mass fraction $f_{\rm ice}$ for possible values of the global dust-to-gas ratio from the canonical $10^{-2}$ down to $10^{-4}$.  The maximum grain size and dust population size distribution is kept constant.  A global dust-to-gas ratio of 10$^{-3.3}$ results in maximum midplane $f_{\rm ice}$ values most consistent with the Galilean satellite bulk compositions and hence is adopted as the reference value throughout this work.

\end{appendix}

\end{document}